\documentclass[a4paper,fleqn,usenatbib]{mnras}

\usepackage[T1]{fontenc}
\usepackage{ae,aecompl}

\usepackage{graphicx}	% Including figure files
\usepackage{amsmath}	% Advanced maths commands
\usepackage{amssymb}	% Extra maths symbols
\usepackage{widetext}
\usepackage{soul} % Using strike out text with st

\def\xx{\boldsymbol{x}}

\def\qq{\boldsymbol{q}}

\def\vv{\boldsymbol{v}}
\def\ppsi{\boldsymbol{\psi}}
\def\dif{\mathrm{d}}
\def\kF{k_{\rm F}}

\title[Cosmology rescaling with massive neutrinos]{How to add massive neutrinos to your $\Lambda$CDM simulation -- extending cosmology rescaling algorithms}

\author[M. Zennaro et al.]{
Matteo Zennaro,$^{1}$\thanks{E-mail: matteo\_zennaro001@ehu.eus}
Ra\'ul E. Angulo,$^{1,2}$
Giovanni Aric\`o,$^{1}$
Sergio Contreras,$^{1}$
\newauthor{} Marcos Pellejero-Ib\'a\~nez$^{1}$
\\
$^{1}$Donostia International Physics Center (DIPC), Paseo Manuel de Lardizabal, 4, 20018 Donostia-San Sebasti\'an, Spain\\
${2}$ IKERBASQUE, Basque Foundation for Science, E-48013, Bilbao, Spain.
}

\date{Accepted XXX. Received YYY; in original form ZZZ}

\pubyear{2019}

\begin{document}
\label{firstpage}
\pagerange{\pageref{firstpage}--\pageref{lastpage}}
\maketitle

% Abstract of the paper (max 250 words)
\begin{abstract}
%The cosmology rescaling algorithm defines a set of time and space transformations that, once applied to the outputs of a cosmological simulation run assuming a given cosmology, allows for reproducing the outputs of another simulation, run in a target cosmology, different from the original one.
%One of its most interesting applications is to Bayesian analysis of cosmological datasets, which, instead of being compared to models, are compared directly to simulations, thus enabling the retrieval of information from unprecedentedly small scales.
Providing accurate predictions for the spatial distribution of matter and luminous tracers in the presence of massive neutrinos is an important task, given the imminent arrival of highly accurate large-scale structure observations.
In this work, we address this challenge by extending cosmology-rescaling algorithms to massive neutrino cosmologies. In this way, a $\Lambda$CDM simulation can be modified to provide nonlinear structure formation predictions in the presence a hot component of arbitrary mass, and, if desired, to include non-gravitational modifications to the clustering of matter on large scales.
We test the accuracy of the method by comparing its predictions to a suite of simulations carried out explicitly including a neutrino component in its evolution equations. We find that, for neutrino masses in the range $M_\nu \in [0.06, 0.3] ~ \mathrm{eV}$ the matter power spectrum is recovered to better than $1\%$ on all scales $k<2~h~\mathrm{Mpc}^{-1}$.
Similarly, the halo mass function is predicted at a few percent level over the range $M_{\rm halo} \in [10^{12}, 10^{15}] ~ h^{-1} ~ \mathrm{M}_{\odot}$, and so do also the multipoles of the galaxy 2-point correlation function in redshift space over $r \in [0.1, 200] ~ h^{-1} ~ \mathrm{Mpc}$. We provide parametric forms for the necessary transformations, as a function of $\Omega_{\rm m}$ and $\Omega_{\nu}$ for various target redshifts.
\end{abstract}

% Select between one and six entries from the list of approved keywords.
% Don't make up new ones.

%%%%%%%%%%%%%%%%%%%%%%%%%%%%%%%%%%%%%%%%%%%%%%%%%%

%%%%%%%%%%%%%%%%% BODY OF PAPER %%%%%%%%%%%%%%%%%%

\section{Introduction}

Cosmology is at the verge of a new era of spectroscopic and photometric data, with upcoming galaxy surveys such as Euclid, DESI, DES, JPAS, or LSST. These data will provide incredible thrust to better constrain the parameters of the cosmological model. At the same time, any deviation from it would indicate the emergence of new physics, thus gauging the room available for extending our models. Among the numerous, possible extensions to the simplest $\Lambda$CDM model, massive neutrinos occupy a privileged spot due to the strong evidence of their existence.

As a matter of fact, observations that neutrinos oscillate among their flavours set a lower bound to their mass: $M_{\nu} = \sum_i m_{\nu, i} \gtrsim 0.06 ~ \mathrm{eV}$ at 95\% confidence level \citep{Gonzalez-GarciaEtal2012, Gonzalez-GarciaMaltoniSchwetz2014, ForeroTortolaValle2014,EstebanEtal2017}. At the same time, the study of the endpoint energy of electrons produced in $\beta$-decay experiments places an upper bound to the neutrino mass, such as $m(\nu_e) < 2.2 ~ \mathrm{eV}$ at 95\% \citep{KrausEtal2005}. Future, planned experiments \citep[such as KATRIN and PTOLEMY,][]{BonnEtal2011,PTOLEMYEtal2019} are expected to increase the sensitivity at least by one order of magnitude. Complementing these efforts, cosmology is already able to produce competitive neutrino mass constraints by exploiting the modifications they induced on the large-scale structure (LSS) of the universe.

For example, from the sole analysis of the CMB temperature anisotropy power spectrum, the total neutrino mass can be constrained to be $M_\nu < 0.537$ eV, at 95\% credibility level, which decreases to $M_\nu < 0.257$ eV at 95\% when considering also CMB polarization signal \citep{PlanckResults2013,PlanckResults2015, PlanckResults2018}.
Combining this information and that coming from clustering measurements, one can tighten such constraints, for example obtaining $M_\nu < 0.28$ eV at 95\% level, from the joint analysis of the likelihoods of CMB, BAO peak position and smoothed clustering \citep{ZennaroEtal2018}, or even $M_\nu < 0.22$ eV  combining CMB data and BOSS clustering using the so called double-probe approach \citep{Pellejero-IbanezEtal2016}.
Even more stringent constraints come from the analysis of the 1D power spectrum of distant quasars (of which we can see the Ly-$\alpha$ forest), giving $M_\nu < 0.12$ eV \citep{Palanque-DelabrouilleEtal2015}. A similar constraint of  $M_\nu < 0.12$ eV at 95\% level has also been obtained by combining CMB and BAO data, including the high-$\ell$ polarization data of Planck \cite{VagnozziEtal2017}. Finally, there are even hints of detection from the combination of abundance and clustering of galaxy clusters \citep{EmamiEtal2017}.

% These results motivate us to pursue an additional increase in the cosmological constraining power with respect to the neutrino total mass, which we believe can be achieved by starting to exploit the information encompassed in the small nonlinear scales. Therefore the scope of this work is to extend the applicability of cosmology rescaling algorithms to massive neutrino cosmologies. This must be done taking into account the peculiar scale dependent growth of matter fluctuations in such cosmologies, as well as a number of subtleties linked to the definition of mass in this context.

Thanks to future surveys, we expect that the value of the total neutrino mass and, perhaps, their mass ordering, will be measured from LSS data. Therefore, making accurate predictions for the spatial distribution of matter and luminous tracers in the presence of massive neutrinos is, at this point, crucial.

By including neutrinos in the solution of Einstein-Boltzmann equations, it is possible to obtain predictions of the linear power spectra of matter in the presence of neutrinos, from codes such as \texttt{class} \citep{Lesgourgues2011} or \texttt{camb} \citep{LewisChallinorLasenby2000}. The linear density solutions, however, are not enough to describe the spatial distribution of matter at small scales and the clustering properties of collapsed objects and luminous tracers.

To this end, different approaches have been proposed and developed in the literature. On the one hand, one can extend perturbation-theory arguments to cosmologies that include massive neutrinos, as, for example, in \cite{BlasEtal2014,FuhrerWong2015}, or develop Effective Field Theories accounting for a hot massive component \citep{SenatoreZaldarriaga2017,deBelsunceSenatore2019}.
On the other hand, one can directly run full $N$-body simulations including a massive neutrino component. There are a number of possible implementations of neutrinos in cosmological simulations: neutrino perturbations can be solved linearly on a 3D grid, as in the \textit{grid-based} method proposed by \cite{BrandbygeHannestad2009}, or they can be traced by directly sampling the distribution function with additional $N$-body particles, as in the \textit{particle-based} method used by \cite{BrandbygeEtal2008, VielHaehneltSpringel2010, Villaescusa-NavarroEtal2013, CastorinaEtal2015, CarbonePetkovaDolag2016}, or even solved on a grid employing a linear response function that is sensitive to the nonlinearities of the cold perturbations, as in the method of \cite{AliHaimoudBird2013}.
Other methods are possible, in particular a hybrid method between the grid- and particle-based ones, proposed by \cite{BrandbygeEtal2010}, or a hybrid between the linear response and the particle-based method proposed by \cite{BirdEtal2018}. Finally, a method combining particle and fluid techniques has been used by \cite{BanerjeeDalal2016}.

Such large amount of attention drawn to neutrino cosmologies has made possible studying in detail many aspects of the matter distribution \citep{BirdViel2012}, halo and void clustering and abundance \cite{CastorinaEtal2014, CastorinaEtal2015, MassaraEtal2015, RaccanelliVerdeVillaescusa-Navarro2019, VagnozziEtal2018, SchusterEtal2019}, halo model \citep{MassaraVillaescusa-NavarroViel2014}, and redshift space distortions \citep{Villaescusa-NavarroEtal2018a, BelEtal2019}.
Nonetheless, comparing LSS observations to predictions remains a difficult task, especially when dealing with biased tracers and when accounting for degeneracies with other cosmological parameters, owing to the huge amount of computational resources needed to simulate the evolution of neutrino universes.

One approach is to develop emulators of the bias of dark matter tracers in the presence of neutrinos.
This approach has been presented in \cite{ValcinEtal2019}, who calibrated an emulator on $\Lambda$CDM simulations that provides predictions for the halo bias in cosmologies with arbitrary neutrino mass, provided that they share the same remaining cosmological parameters (most importantly the value of $\sigma_8$, the linear amplitude of matter fluctuations at scale $R=8~h^{-1}~\mathrm{Mpc}$).

In this work we propose an alternative method to make predictions for neutrino cosmologies, with emphasis on numerical efficiency and accuracy. Specifically, we extend the simulation rescaling algorithm presented in \cite{AnguloWhite2010} to cosmologies with massive neutrinos. This technique allows to quickly create 3D nonlinear mass density and velocity fields, including halos and their substructures, at an almost negligible computational cost.
The method has already been tested and applied in cases that did not include any non-cold component, finding that it works with very high accuracy \citep[e.g.][]{AnguloWhite2010, RuizEtal2011, MeadEtal2015, AnguloHilbert2015, RennebyHilbertAngulo2018}.

A distinctive advantage of our approach is that any computational effort used in the production of a given high-quality neutrinoless $\Lambda$CDM simulation is reused in the predictions for massive neutrino scenarios; explicitly, these predictions will inherit the quality and realism of the original simulation in terms of nonlinear growth of structures, internal properties of haloes, velocity bias or assembly bias. This enables a realistic and sophisticated modelling of the galaxy population and thus allowing an accurate way to forecast and interpret the wealth of upcoming LSS data.

This paper is organized as follows. In Sec. \ref{sec::neutrinos} we review the relevant aspects of neutrino cosmology, focusing on the quantities we seek to reproduce through the cosmology scaling. In Sec. \ref{sec::method}, after recalling the main aspects of the cosmology rescaling algorithm, we specify the modifications needed to account for massive neutrinos. We test such extended algorithm by quantifying departures in the linear variance and matter power spectrum in Sec. \ref{sec::theoryscale}. Finally, in Sec. \ref{sec::sims} we perform comparisons between neutrinoless $\Lambda$CDM simulations scaled to include massive neutrinos, and simulations actually run including massive neutrinos in their evolution. In Sec. \ref{sec::conclusion}, we present our conclusions.

\section{Massive neutrinos}\label{sec::neutrinos}
In this section, we review the main effects massive neutrinos induce on the evolution of the background and of the perturbations in the Universe. We do so both from an analytic standpoint, by adopting a fluid description of the cosmological neutrino distribution, and from a numerical perspective, by introducing a set of N-body simulations that incorporate a massive neutrino component. Please note that we focus on the aspects that are either useful for our implementation of neutrinos or that we want to reproduce in simulations. For a more general treatment of the effects of neutrinos in cosmology we refer the reader to the comprehensive work of \cite{LesgourguesPastor2006,LesgourguesPastor2012,LesgourguesPastor2014}.

We parametrize the total matter content of the Universe as the sum of a cold and hot component,
\begin{equation}
  \Omega_{\rm m} = \Omega_{\rm cold} + \Omega_{\nu},
\end{equation}

\noindent where the cold component is given by the sum of the baryon and cold dark matter density parameters ($\Omega_{\rm b}$ and $\Omega_{\rm cdm}$, respectively) and the hot component is given by the neutrino density parameter. The remaining energy budget of the Universe is split among the contributions of radiation $\Omega_{\rm r}$, dark energy $\Omega_{\rm de}$ and curvature $\Omega_{\rm k}$. In this work we consider the case of three degenerate massive neutrinos, and we refer to the total neutrino mass as the sum of the masses of the single species, $M_\nu = \sum_i m_{\nu,i}$. Moreover, we will always work in the context of a flat universe, therefore assuming $\Omega_{\rm k} = 0$ in all cases.

Neutrinos become non-relativistic quite late in the thermal history of the Universe, their non-relativistic transition taking place at $1 + z_{\rm nr} \simeq 1890 ~ (m_\nu/1 ~ \mathrm{eV})$. This is much later than the time at which they decouple, i.e. when their interaction rate falls below the expansion rate of the Universe (roughly at $z_{\rm dec} \simeq 10^9$, when the background temperature is roughly $1 ~ \mathrm{MeV}$). For this reason, their momentum distribution remains frozen to that of a relativistic Fermi-Dirac species, even when they are already non-relativistic. The contribution of massive neutrinos to the Hubble function follows directly from the integration of their momentum distribution, which controls the time-dependent evolution of the neutrino density. Their density evolves as that of the radiation component of the Universe, proportionally to the expansion factor to the fourth power, until their non relativistic transition. Afterwards, it asymptotically tends to evolve as matter does, that is proportionally to the expansion factor to the third power.

We also note that the radiation content, in cosmologies without massive neutrinos, is usually given by the sum of the photon and massless neutrino contributions. In a more general context, this is given instead by
\begin{equation}
  \Omega_{\rm r} = \Omega_{\gamma} \left[ 1 + \Delta N_{\rm eff} \dfrac{7}{8} \left( \dfrac{4}{11} \right)^{4/3} \right],
  \label{eq::omega-rad}
\end{equation}
where $\Delta N_{\rm eff}$ is the corrected number of relativistic species, chosen to guarantee that the effective number of relativistic species in the early universe is $N_{\rm eff} = 3.046$, i.e. what is expected for three neutrinos undergoing non-instantaneous decoupling \citep{ManganoEtal2005}.

One of the consequences of the freezing of the neutrino momentum distribution is that they are characterized, throughout the thermal history of the Universe, by relatively large velocities compared to any other massive particle. Picturing neutrinos as a fluid \citep{ShojiKomatsu2010}, such velocities act as an effective pressure term and set a scale below which the growth of neutrino perturbations is strongly suppressed. Such scale is called the \textit{free-streaming scale} and, in fluid approximations, can be written as
\begin{equation}
  k_{\rm fs} = \sqrt{\dfrac{3}{2} \Omega_{\rm m}(a)} \dfrac{a H(a)}{c_{\nu}(a)}.
\end{equation}
We have introduced here the speed of sound corresponding to the effective neutrino pressure \citep{BlasEtal2014},
\begin{equation}
  c_{\nu}(a) = \dfrac{\delta P_\nu}{\delta \rho_\nu} \simeq \dfrac{123.423}{a ~ m_\nu} ~ \mathrm{km} ~ \mathrm{s}^{-1},
  \label{eq::csound}
\end{equation}
where $m_\nu$ is expressed in electronvolts.

Not only are neutrino perturbations affected by such suppression, but, by solving the joint equations of growth of cold and hot matter, the neutrino pressure term induces a scale dependent suppression of the growth also for the cold components.

\subsubsection{Clustering}

\begin{figure}
  \includegraphics[width=0.48\textwidth]{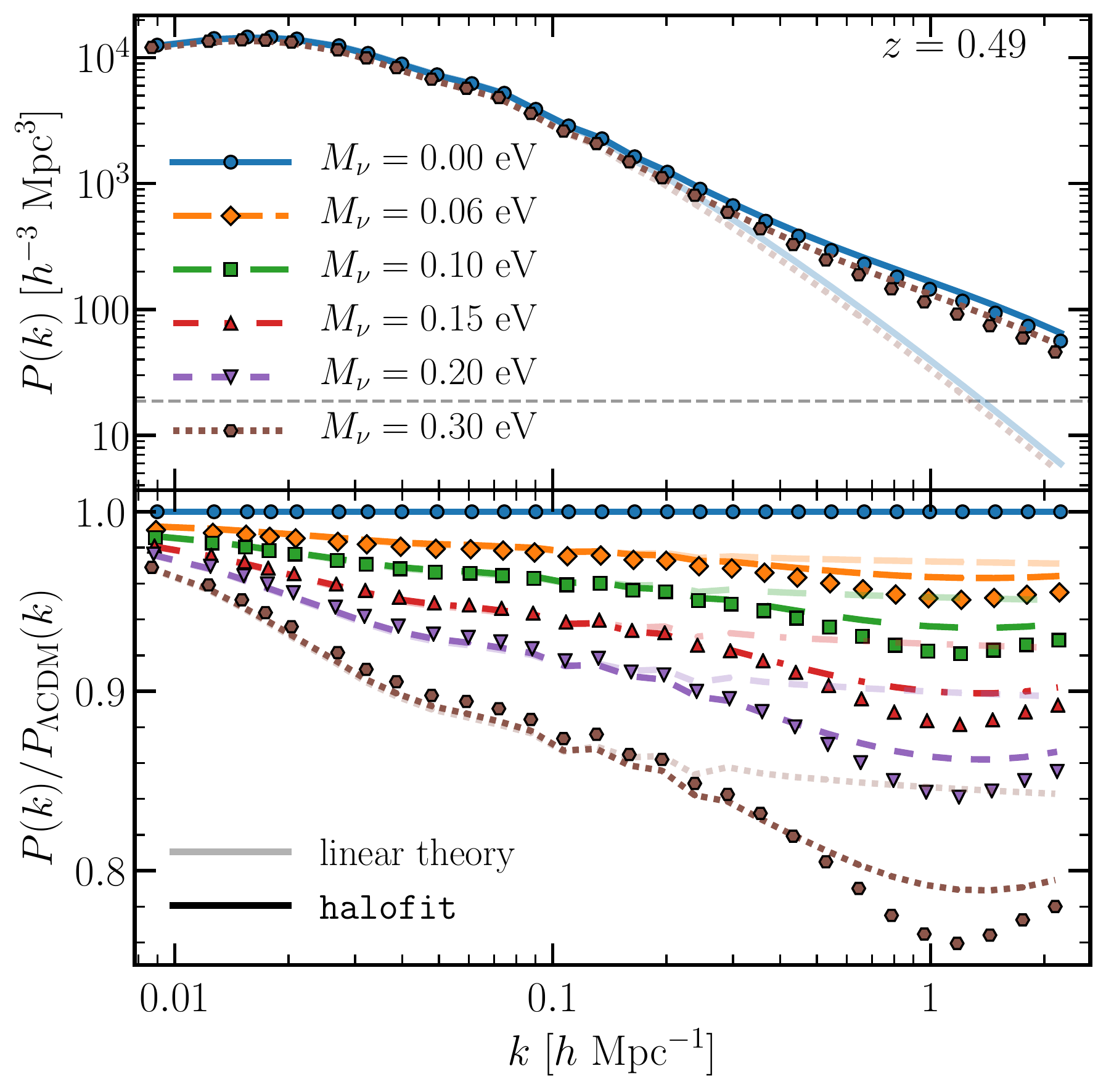}
  \caption{The cold matter (cdm + baryons) power spectrum in the presence of massive neutrinos. Points are measurements in the simulations, lines are predictions in linear theory (transparent colours) and applying \texttt{halofit} (non-transparent colours). The simulations employed have different neutrino masses, but same baseline cosmology, and share the same large scale normalization. Here results at redshift $z \simeq 0.5$ are shown. The upper panel shows the power spectra superimposed to their corresponding theoretical predictions; the grey, horizontal line marks the level of the shot-noise. The bottom panel shows the ratio between the measurements in the simulations and the predictions. The suppression due to neutrino free streaming is larger for larger neutrino masses. Moreover, neutrinos induce a peculiar spoon-like feature in the non linear matter power spectrum at relatively small scales ($k \simeq 1 ~ h ~ \mathrm{Mpc}^{-1}$).}
  \label{fig::neutrino-pk}
\end{figure}

In Fig. \ref{fig::neutrino-pk} we show the effect of massive neutrinos on the two point statistics of the mass field. We do so by studying a set of neutrino simulations whose technical aspects will be detailed in Sec. \ref{sec::sims}. We analyse the cold matter power spectrum computed in cosmologies with massive neutrinos, with increasing neutrino total mass $M_\nu = \{0.06, 0.1, 0.15, 0.2, 0.3\} ~ \mathrm{eV}$ and compare it to a $\Lambda$CDM case with $M_\nu = 0$ eV. At high wavelenghts, the neutrino induced suppression of clustering is $\Delta P/P \sim 8 \, \Omega_{\nu}/\Omega_{\rm m}$ -- spanning from $\sim 4\%$ at $k = 1 ~ h ~ \mathrm{Mpc}^{-1}$ for the lightest neutrino mass, to $\sim 25\%$ at the same scale in the case of the heaviest mass.

The evolution of the cold matter power spectrum in the nonlinear regime exhibits a well known, spoon like shape, \citep[see, for example,][]{BirdVielHaehnelt2012}. The reason resides in the combination of the suppression of the matter clustering induced by neutrinos, and the fact that neutrinos move the nonlinear scale $k_{\rm nl}$ to larger modes. Alongside our simulated results, we show nonlinear predictions obtained by applying the \texttt{halofit} algorithm \citep{TakahashiEtal2012} to the cold linear power spectrum predicted by \texttt{class}.

\subsubsection{Halo abundance}

\begin{figure}
  \includegraphics[width=0.48\textwidth]{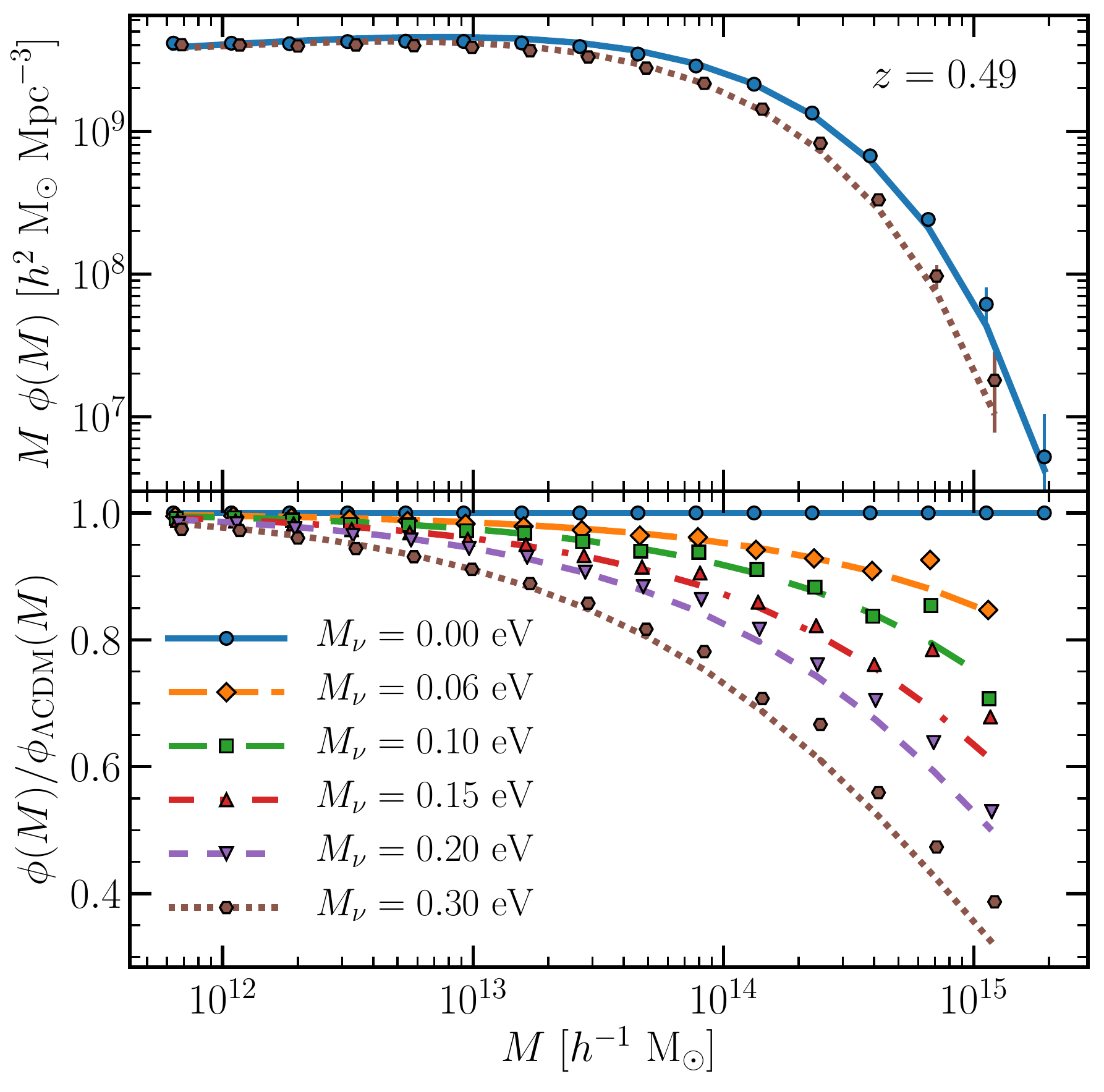}
  \caption{The logarithmic halo mass function, $\phi(M) = \dif n / \dif \ln M$, multiplied by $M$ to reduce the dynamical range of the y axis. We show for each mass bin the associated Poisson error. The upper panel shows the two extreme cases of $M_\nu=0$ and $0.3$ eV. The lower panel shows, for six choices of neutrino mass spanning the above range, the ratio between the halo mass function in the presence of neutrinos and its corresponding $\Lambda$CDM case. Lines are theoretical predictions obtained as detailed in the text, while points are measurements in our simulations. In the presence of massive neutrinos, massive halos become rarer, when compared to halos of the same mass in pure $\Lambda$CDM cosmologies. As a consequence, the high mass end of the halo mass function is suppressed, the suppression being larger for larger neutrino masses.}
  \label{fig::neutrino-hmf}
\end{figure}

We present in Fig. \ref{fig::neutrino-hmf} the halo mass function in the neutrino simulations, compared to the $\Lambda$CDM case. We compute the quantity $\dif n / \dif \ln M$, considering as halo mass the quantity $M_{200}$, that is the mass enclosed in a sphere whose average density is equal to 200 times the comoving critical density of the Universe.

To model the halo mass function in the presence of neutrinos, we use the fits obtained by \cite{AnguloEtal2012} without massive neutrinos, being careful to compute the variance of the matter distribution only in terms of the cold component, $\sigma_{\rm cold}$, \citep{CastorinaEtal2014,CastorinaEtal2015}. This is justified by the fact that the neutrino free-streaming scale is, at all times, significantly larger than the typical size of the most massive haloes. As a consequence, the dynamics of the collapse are dominated by the cold dark matter and baryons \citep[see, for example,][who showed that the collapse threshold changes by less than $0.1\%$ in the presence of neutrinos]{IchikiTakada2012}.

It comes therefore as no surprise that the main effect of neutrinos is to suppress the high mass end of the halo mass function. As a matter of fact, for the same $\Omega_{\rm m}$, haloes of a given mass are significantly rarer (since there is \textit{effectively} less matter collapsing). As shown in Fig. \ref{fig::neutrino-hmf}, this can result in finding from $\sim 20\%$ to $\sim 60\%$ less haloes in the high mass bin centered in $10^{15} ~ h^{-1} ~ \mathrm{M}_{\odot}$.

These effects are specific signatures of massive neutrinos and are dependent on the value of the neutrino total mass parameter. Therefore, they provide us with a way to constrain the neutrino total mass through cosmological observations. We aim at reproducing such effects in neutrino simulations that, instead of being run in actual neutrino cosmologies, can be obtained by rescaling a $\Lambda$CDM simulation, originally run without any hot massive component.

It has been shown by \cite{AnguloWhite2010}, AW10 hereafter, that structure formation can be predicted by matching the linear density amplitude $\sigma(M)$. The fact that the neutrino induced effects can be captured by $\sigma_{\rm cold}$ leads us to postulate the Anstatz that neutrino simulations can be accurately reproduced by extending the ideas of AW10. We will develop and test this idea further in the next section.

\section{Method: extending cosmology rescaling to massive neutrino cosmologies}\label{sec::method}
In this section we will review the scaling method presented in \cite{AnguloWhite2010} and extend it to deal with the presence of massive neutrinos. Note that we will meticulously follow the philosophy of the original paper, stressing the modifications necessary to model massive neutrinos. The goal of the procedure is to find the transformation of times and lengths such that a simulation that was run assuming a given cosmology (that we will refer to as \textit{original cosmology} throughout this paper) resembles as closely as possible a simulation run in a different cosmology (that we will refer to as \textit{target cosmology}). Such resemblance can be quantified in terms of matter clustering, halo abundance and bias and galaxy clustering in redshift space.

In our case study, we will always consider the rescaling of a $\Lambda$CDM cosmology (original) so that it reproduces a neutrino cosmology (target). However, note that it is also possible to consider scaling a neutrino cosmology to a different value of $\Omega_{\nu}$.

The general idea of the scaling algorithm can be summarised in a few steps:

\begin{itemize}
  \item determine the transformation of lengths ($s$) and time ($z^*$) that minimizes the difference between the linear amplitude of perturbations in the original and target cosmology;
  \item select the simulation output in the original cosmology at the closest redshift to the one identified through the minimization, $z^*$;
  \item apply the length transformation to the positions of particles/haloes/subhaloes in this snapshot, which corresponds now to a simulations with different volume $(sL_{\rm box})^3$ and particle mass ($s^3 \Omega_{\rm m}^\prime M_{\rm part} / \Omega_{\rm m}$, where primed refers to the target cosmology);
  \item correct the large scales modes, to account for the different bulk motions between original and target cosmologies;
  \item correct nonlinear velocities;
  \item correct halo concentrations.
\end{itemize}
We will now go through these steps underlying where we need to account for the presence of massive neutrinos.

\subsection{Relevant modifications to include massive neutrinos}

In each cosmology, the typical level of inhomogeneity associated to a scale $R$ is given by the linear variance of the density field smoothed on that scale. In Press-Schechter theory, this uniquely determines the bias and halo mass function. For this reason, the first step of the procedure is to try to match, as closely as possible, the linear amplitude of the matter density field of the target cosmology.

More specifically, we fix the redshift at which we want to reproduce the target cosmology, $z_t$, and we assume we have run the original simulation in a cosmological box of comoving side $L_{\rm box}$. In order to mimic the target cosmology at redshift $z_t$ we will need to consider the original cosmology at a different redshift, $z_*$. Lengths will also be different and we will need to modify the simulation run in the original cosmology so that it has box size $L^\prime_{\rm box} = s L_{\rm box}$. The modified simulation will mimic a simulation run in the target cosmology with box size $L^\prime_{\rm box}$ at $z_t$. We find the scaled redshift and the length scale factor $s$ by minimizing the difference between the linear amplitude of fluctuations $\sigma(R)$ in the original and target cosmologies over a given interval of scales, $R \in [R_1, R_2]$,
\begin{equation}
  \delta^2_{\rm rms}(s, z_*) \equiv \dfrac{1}{\ln(R_2/R_1)} \int _{R_1}^{R_2} \dfrac{\dif R}{R} \left[ 1 - \dfrac{\sigma(s^{-1} R, z_*)}{\sigma^\prime(R, z_{\rm t})} \right]^2.
\label{eq::minimization}
\end{equation}

When including massive neutrinos, as discussed in Sec. \ref{sec::neutrinos}, the physical meaning of the linear amplitude of matter fluctuations is preserved provided that we substitute the total matter density field with the cold matter (cold dark matter + baryons) density field. Note that this is a direct consequence of neutrinos being characterized by a velocity dispersion that is, at all redshifts, comparable or larger than typical velocity dispersions in haloes, thus preventing the former from playing a significant role in the collapse process of the latter. We therefore modify $\sigma$ in Eq. \ref{eq::minimization} from its version in AW10, which now reads
\begin{equation}
  \sigma^2(R,z) = \int_0^{\infty} \dfrac{k^2 \dif k}{2 \pi^2} W^2(kR) D_{\rm cold}^2(k,z) P_0(k),
\end{equation}
where $W(x) \equiv 3 (\sin x - x \cos x) / x^3$ is the Fourier transform of the top-hat window function, $P_0$ is the total matter power spectrum at redshift zero and $D_{\rm cold}(k, z)$ is the growth factor (in Fourier space) for the cold matter component. Note that, in the presence of massive neutrinos, it depends on both scale and time, and, therefore, cannot be factorized out of the integral. The cold growth factor, in our convention, is always defined with respect to the total matter power spectrum, so that for any species considered we have $P_i(k,z) = D_i^2(k,z) P_0(k)$.

To find the scale dependent cold matter growth factor, we use the code \texttt{reps} \citep{ZennaroEtal2016}, which employs a three-fluid description, comprising a pressureless cold dark matter fluid, a presureless baryon fluid and a neutrino fluid characterized by the effective speed of sound presented in Eq. \ref{eq::csound}.
We therefore solve the set of linearised equations of growth of fluctuations for the three components. The pressure term introduced for neutrinos provides the correct redshift- and scale-dependent suppression for the cold matter growth factor. Note that neglecting the pressure term associated to the baryon fluid is in principle not justified.
However, in this work we are interested in characterizing neutrino effects, which have been shown to be separable from those of baryons \citep{MummeryEtal2017}.

\begin{figure*}
  \includegraphics[width=0.98\textwidth]{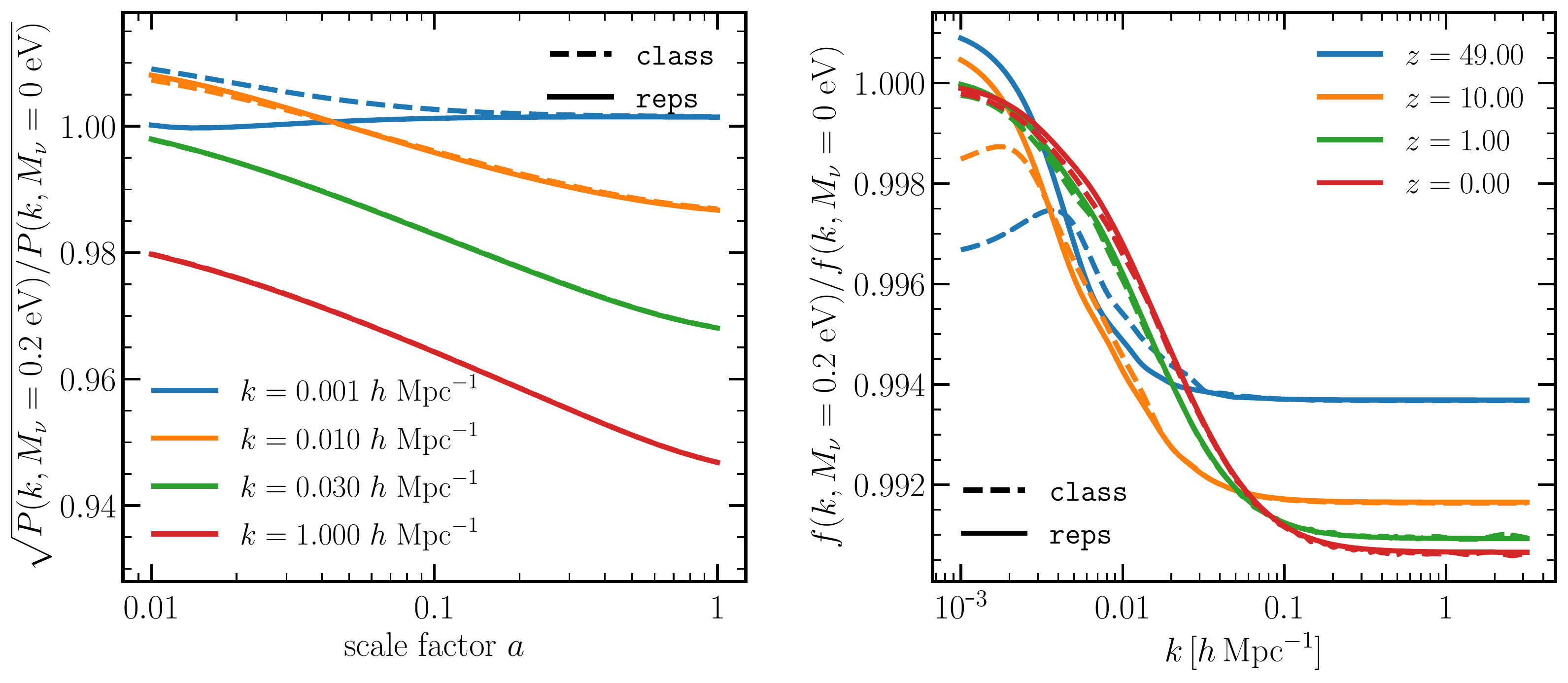}
  \caption{\textit{Left:} the ratio of the scale dependent square roots of the power spectra with and without neutrinos as computed using \texttt{reps} (solid lines) and \texttt{class} (dashed lines). The solutions for four wavemodes are shown as a function of the scale time. \textit{Right:} the ratio of the scale-dependent growth rates with and without neutrinos, computed with \texttt{reps} (solid lines) and \texttt{class} (dashed lines). The neutrino mass arbitrarily chosen for this example is $M_\nu=0.2$ eV.}
  \label{fig::nu-scaledep}
\end{figure*}

In Fig. \ref{fig::nu-scaledep} we show an example of the scale dependent growth factor and growth rate, both as obtained using \texttt{reps} and \texttt{class}. The former provides purely Newtonian solutions, while the latter solves the Einstein-Boltzmann equations including non-gravitational coupling such as that of radiation and baryons on scales approaching the cosmological horizon.
Part of the discrepancy among the codes observable on large scales is due to such effect \citep[see, for example,][]{BrandbygeEtal2016}, while part is due to the fact that, while the results of \texttt{class} used here are in synchronous gauge, the results of \texttt{reps} are purely Newtonian (and could be interpreted in General Relativity assuming some \textit{ad hoc} gauge, different from the synchronous one).
Since the outputs of \texttt{reps} are designed to agree with the dynamics of a Newtonian simulation, we will use this code in order to make predictions for the (cold) matter power spectrum, growth factor and growth rate in the rest of this work. However, we will also present a method to reproduce, if desired, these large scale contributions to the power spectrum.

From Fig. \ref{fig::nu-scaledep} we can appreciate that the suppression of the clustering due to massive neutrinos evolves in time differently for different wavemodes. In particular, large scales, such as $k \simeq 10^{-3} ~ h ~ \mathrm{Mpc}^{-1}$, evolve very little, while small scales, such as $k \simeq 1 ~ h ~ \mathrm{Mpc}^{-1}$, go from showing a suppression with respect to the $\Lambda$CDM case of roughly 2\% at high redshift, to almost 6\% at low redshift.

Note that velocities in the scaled simulation need to be corrected to comply with the different growth history of the target cosmology. Following once again AW10 this can be achieved through
\begin{equation}
  \vv^{\prime} = \dfrac{a^{\prime} f^{\prime}(a^{\prime}) H^{\prime}(a^{\prime})}{a f(a) H(a)} ~ \vv,
  \label{eq::vel-corr}
\end{equation}
where $f \equiv \dif \ln D/  \dif \ln a$ is the growth rate, primes quantities are in the scaled simulation and non-primed ones in the original cosmology. In the case of neutrino cosmologies the growth rate is scale dependent, $f = f(k, z)$. However, as shown in Fig. \ref{fig::nu-scaledep}, it exhibits two asymptotic regimes at large and small scales. We choose to use in Eq. \ref{eq::vel-corr} the value at small scales (obtained at the arbitrarily fixed wavemode $k=0.5 ~ h ~ \mathrm{Mpc}^{-1}$). This choice accurately corrects the velocity of the underlying bulk motions of overdensities on nonlinear scales, but is not accurate for those on larger scales. We will address in the next paragraphs residual corrections that further increase the accuracy of the rescaling procedure.

First, even though the matching of the linear amplitude of fluctuations is tailored to provide a good agreement of the clustering of the scaled and target simulations on small scales, because it is dominated by the scales that contribute the most to the overall inhomogeneity, residual inaccuracies might affect the large scale clustering and bulk motion.
However, the evolution of the large scales is easily modelled in linear theory. The solution is, therefore, to correct the positions and velocities of the particles (or haloes, or subhaloes) of the scaled simulation.
This is done to greater accuracy in Lagrangian space. Once the corresponding Lagrangian position of each object is found, we apply a displacement field such that the large scale fluctuations of the original cosmology are subtracted. We then proceed with the rescaling procedure, and, at the end, we apply another displacement to add the contribution of the long waves of the target cosmology.

In terms of displacement fields, we can therefore move our particles so that
\begin{equation}
  \xx = \qq - \ppsi_{\rm o}(\qq) + \ppsi_{\rm t}(\qq),
  \label{eq::lss-correction}
\end{equation}

\noindent where $\qq$ are the Lagrangian positions obtained iteratively with the method described in AW10, $\ppsi_{\rm o}$ and $\ppsi_{\rm t}$ are the displacement fields corresponding to the original and target cosmologies, respectively. Note that we only employ modes smaller than the nonlinear scale $k_{\rm nl} = 1 / R_{\rm nl}$, defined so that $\sigma(R_{\rm nl}) = 1$.

A second consideration regards the coupling, mentioned above, of radiation and baryon perturbations on large scales, an effect that is neglected in Newtonian N-body simulations. If the simulation is started (as in our case) with an initial power spectrum obtained scaling the $z=0$ spectrum back in time, then the matter power spectrum measured in the simulation (on large, linear scales) will match the solutions of the Einstein-Boltzmann equations only at low redshift.
At higher redshifts an unavoidable mismatch will appear on the scales approaching the cosmological horizon. If desired, this effect can be reintroduced, by incorporating it in the large scale correction presented in Eq. \ref{eq::lss-correction}. The target displacement field will therefore contain a double contribution
\begin{equation}
  \ppsi_{\rm t} = \ppsi^{GR}_{\rm t} + \ppsi^{NG}_{\rm t},
\end{equation}
where the term $\ppsi^{GR}_{\rm t}$ corrects the difference in the large scale matter overdensities due to the different expansion and background gravitational potential between the original and the target cosmology (i.e. the gravitationally induced differences), while the term $\ppsi^{NG}_{\rm t}$ accounts for these non-gravitational couplings.

Moreover, an additional correction can be applied to velocities of particles that reside in haloes, $\vv_{\rm in-halo}$, to account for any residual nonlinear effect. We modify the correction presented in AW10, replacing the total matter density parameter $\Omega_{\rm m}$ with the cold matter one, $\Omega_{\rm cold}$, thus obtaining
\begin{equation}
  \vv^\prime_{\rm in-halo} = \sqrt{\dfrac{a ~ \Omega_{\rm cold}^\prime}{a^\prime ~ \Omega_{\rm cold}}} ~ \dfrac{h^\prime}{h} ~ s ~ \vv_{\rm in-halo},
\end{equation}
where primed quantities are in the target cosmology and $s$ is the length scale factor and $h = H_0/100$ the Hubble parameter.

Finally, even better agreement of the clustering on scales dominated by the one-halo term can be achieved when accounting for the different concentrations haloes have in the two cosmologies. We expect halo concentrations to be different depending on the different halo formation time. We therefore define a displacement field accounting for the different density profile shapes expected at the time of the original simulation and at the time of the target simulation. We then proceed to apply it to all particles residing in a bound halo, on a halo-by-halo basis.

In summary, the fundamental considerations that we need to keep in mind to scale a $\Lambda$CDM cosmology to a neutrino cosmology can be outlined as:
\begin{itemize}
  \item when finding the scale time and scale length factors, we need to minimize the difference in the amplitude of cold matter linear perturbations, instead of using the one referred to the total matter;
  \item the growth factor, used to correct large scales, is scale dependent (and the scale dependence evolves with redshift);
  \item consequently, also the growth rate, used to correct large scale velocities, is scale and redshift dependent;
\end{itemize}

\subsection{Examples}\label{sec::theoryscale}
\begin{figure*}
  \includegraphics[width=\textwidth]{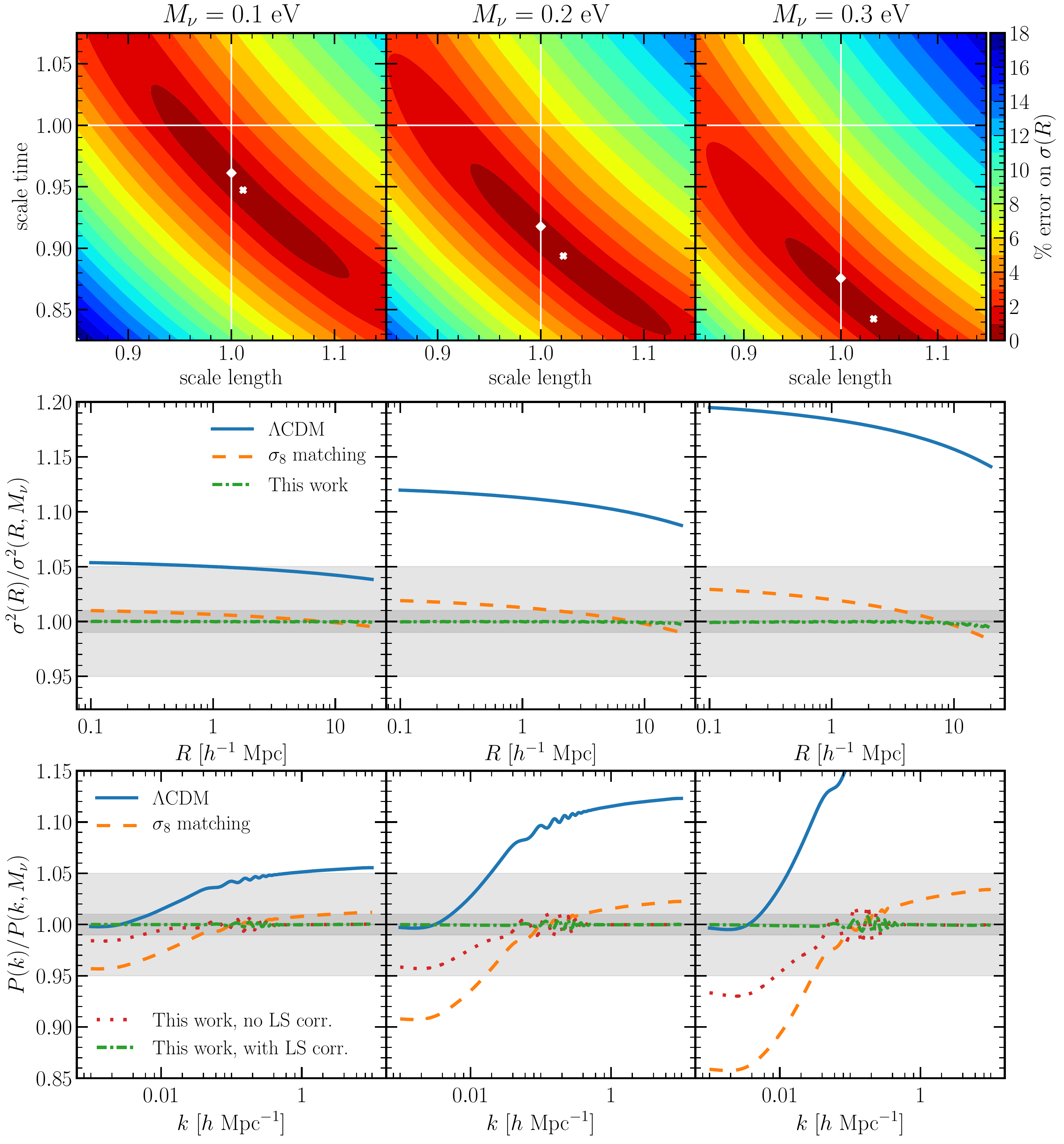}
  \caption{\textit{Top row}: The log-likelihood surface associated to the difference between the variance in the scaled and in the target cosmology. The white crosses mark the minima, i.e. the choices of length and time scaling factors that, applied to the original $\Lambda$CDM cosmology, allow it to best reproduce the variance in the target neutrino cosmology at the target redshift $z_{\rm t} = 0$. The diamonds mark the scaling time selected if we just match the value of $\sigma_8$.
  \textit{Middle row}: The accuracy of the rescaled variance compared to the target one. The light grey and dark grey shaded areas mark the 5\% and 1\% levels, respectively. Blue solid, orange dashed and green dash-dotted lines show, respectively, the variance in the original cosmology at the target redshift, the variance obtained through the $\sigma_8$ matching and the variance of the scaled cosmology (this work), all divided by the variance computed in the target cosmology. The accuracy is at better than 1\% level for all of the smoothing scales considered.
  \textit{Bottom row}: The same ratio, but for the linear power spectrum. Additionally, we show in red dotted lines the ratio for the case in which we do not apply the large scale (LS) correction described in the test. In the case involving the full scaling procedure, the ratio between the scaled and the target linear power spectrum results to be well within the 1\% accuracy level.}
  \label{fig::theory}
\end{figure*}

In the present section we employ the method described up to here, to rescale the linear theory predictions of a cosmology that does not include massive neutrinos, to one with non-null neutrino total mass.
We consider three target cosmologies, with different neutrino masses, $M_\nu = \{0.1, 0.2, 0.3\}$ eV. All cosmologies share the same baseline cosmological parameters and have the same amount of total matter, the cold dark matter density parameter consequently being different, according to
\begin{equation}
  \Omega_{\rm cdm} = \Omega_{\rm cdm}^{\Lambda\mathrm{CDM}} - \Omega_\nu,
\end{equation}
where the neutrino density parameter is linked to the neutrino total mass though $\Omega_\nu = M_\nu /(93.14~h^2)$.

We consider two cases: along with the full minimization of the linear density amplitude in terms of length and time scaling, we also analyse the result of just selecting the time at which there is a matching of the linear amplitude smoothed at a scale of $8 ~ h^{-1} ~ \mathrm{Mpc}$, $\sigma_8$.

An example of such minimization is presented in the top row of Fig. \ref{fig::theory}. In order to reproduce cosmologies with increasingly large neutrino masses, we need to consider the $\Lambda$CDM cosmology at increasingly earlier redshift. This is true both when we just match the value of $\sigma_8$ and when we perform the full minimization. However, the time selected in the two cases is slightly different. With the full minimization we also find that the matching is more accurate if we expand the lengths in the $\Lambda$CDM cosmology.

From the top row of Fig. \ref{fig::theory} we expect that both by matching $\sigma_8$ and by performing the full minimization the linear amplitude in the scaled cosmology and in the target one differ by less than $1\%$. This is confirmed in the middle row of the same figure, where we compare the linear variance of the matter field smoothed with spherical top-hat functions with different radii computed in the scaled $\Lambda$CDM cosmology and in the target neutrino cosmology. The filtering scales span the interval $R \in [0.1, 10] ~ h^{-1} ~ \mathrm{Mpc}$, which covers the range of typical Lagrangian sizes of haloes. We can see that the full minimization ensures an almost perfect agreement on all the scales considered, while, as expected, the $\sigma_8$ matching performs slightly worse the farther we are from $R=8 ~ h^{-1} ~ \mathrm{Mpc}$.

In the bottom row of Fig. \ref{fig::theory} we compare the linear matter power spectrum between the scaled $\Lambda$CDM cosmology and the neutrino cosmology. In this case it is even more apparent that by just matching $\sigma_8$ we introduce a (scale dependent) error going from 5\% to 15 \% for $M_\nu=0.1 ~ \mathrm{eV}$ and $M_\nu=0.3 ~ \mathrm{eV}$ respectively. We notice that even performing the full minimization, although the agreement largely improves on small scales, there is a mismatch at the 2-7\% level on large scales. However, we predict that, with the full minimization procedure and applying a correction of the large, linear scales,  we can match the scaled power spectrum with sub-percent precision.

\section{Testing with simulations}\label{sec::sims}

In the present section we proceed to test our method using N-body simulations. In particular, we will scale a $\Lambda$CDM simulation in order to add massive neutrinos. Each time, we compare the scaled simulation to the corresponding simulation \textit{actually} run including neutrinos. We perform the comparison in terms of matter clustering in real space, halo abundance and galaxy clustering in redshift space.

We simulate the evolution of the matter field in six cosmologies, characterized by different neutrino masses, namely $M_\nu = \{0, 0.06, 0.1, 0.15, 0.2, 0.3\}$ eV. Our simulations are performed with a modified version of the N-body code \texttt{GADGET} \citep{Springel2005}, extensively based on its lean version, \texttt{L-GADGET-3}, presented in \cite{AnguloEtal2012}. To follow the evolution of neutrinos, we included the implementation presented by \cite{AliHaimoudBird2013}.
This method consists in updating the potential mesh with the contribution of neutrino densities, computed solving the Boltzmann equation in Fourier space. Even if  the solution is found linearising for small neutrino overdensities, the total potential is the one computed in the simulation, and accounts for the nonlinear evolution of the cold components.
This approximation, albeit imperfect, works remarkably well  because neutrinos are expected never to develop significant nonlinearities.\footnote{ It is important to note that, however small, the nonlinear evolution of neutrinos could in principle play a role in the nonlinear growth of cold matter, a case that this method would fail to describe. A more complete, though computationally expensive, version of the linear response  method has been presented in \cite{BirdEtal2018}. However, discrepancies are mainly expected to be found at the neutrino power spectrum level, while \cite{BirdEtal2018} found that the cold matter power spectrum obtained either with the method utilized here, with the more accurate hybrid method, or with particle-based simulations differ by less than 0.1\%.}

Our baseline cosmology is described by a flat $\Lambda$CDM model, with total matter density parameter $\Omega_{\rm m} = 0.307112$, baryon density parameter $\Omega_{\rm b} = 0.048252$,  Hubble parameter $H_0 = 67.77 ~ \mathrm{km} ~ \mathrm{s}^{-1} ~ \mathrm{Mpc}^{-1}$, optical depth at recombination $\tau = 0.0952$, spectral index of primordial fluctuations $n_s = 0.9611$ and amplitude of scalar perturbations at the pivotal physical scale $k_{\rm pivot} = 0.05 ~ \mathrm{Mpc}^{-1}$ equal to $A_s = 2 \times 10^{-9}$.
When adding massive neutrinos we keep the total matter density at present time fixed, thus modifying the cold dark matter fraction according to $\Omega_{\rm cdm} = \Omega_{\rm m} - \Omega_{\rm b} - \Omega_\nu$.
The photons have present day temperature $T_\gamma = 2.7255$ K \citep[in agreement with][]{Fixsen2009}, corresponding to $\Omega_\gamma = 5.3758 \times 10^{-5}$. The relativistic neutrino contribution to the radiation is obtained through Eq.(\ref{eq::omega-rad}).

All simulations follow the evolution of $1056^3$ particles of cold matter in a periodic comoving cube of side $L_{\rm box} \simeq 700 ~ h^{-1} ~ \mathrm{Mpc}$ and softening length set to 1/50th of the mean intraparticle distance, $\epsilon \simeq 13 ~ h^{-1} ~ \mathrm{kpc}$. They are initialized at $z_{\rm i} = 49$, using, for each cosmology, the present day power spectrum rescaled back in time at the initial redshift with the method described in \cite{ZennaroEtal2016}. To the end of reducing the importance of cosmic variance, for each cosmology we run two paired simulations, with the same initial fixed amplitudes and phase realizations shifted by $\pi$ radians \citep{AnguloPontzen2016}. We employ a slightly different volume for each simulation, in order to be able to compare them with the rescaled $\Lambda$CDM simulations (which, after applying the rescaling transformations, will have a effectively larger volume). We list in Table \ref{tab::sims} the different volumes used.

\begin{table}
\centering
  \begin{tabular}{c c c c c}
    $M_\nu$ & $\Omega_{\rm cold}$ & $L_{\rm box}$ & $m_{\rm p}$ & $\sigma_8$ \\
    $[\mathrm{eV}]$ &  & $[h^{-1} ~ \mathrm{Mpc}]$ & $[10^{10} \mathrm{M}_{\odot}]$ & \\
    \hline
    0.00 & 0.30711 & 700.0 & 2.482 & 0.800 \\
    0.06 & 0.30571 & 705.2 & 2.527 & 0.790 \\
    0.10 & 0.30477 & 707.9 & 2.548 & 0.783 \\
    0.15 & 0.30361 & 711.6 & 2.577 & 0.773 \\
    0.20 & 0.30244 & 715.3 & 2.608 & 0.763 \\
    0.30 & 0.30010 & 723.6 & 2.677 & 0.742 \\
    \hline
    \end{tabular}
  \caption{The specifications of the set of simulations considered.}
  \label{tab::sims}
\end{table}

\begin{figure*}
  \includegraphics[width=0.48\textwidth]{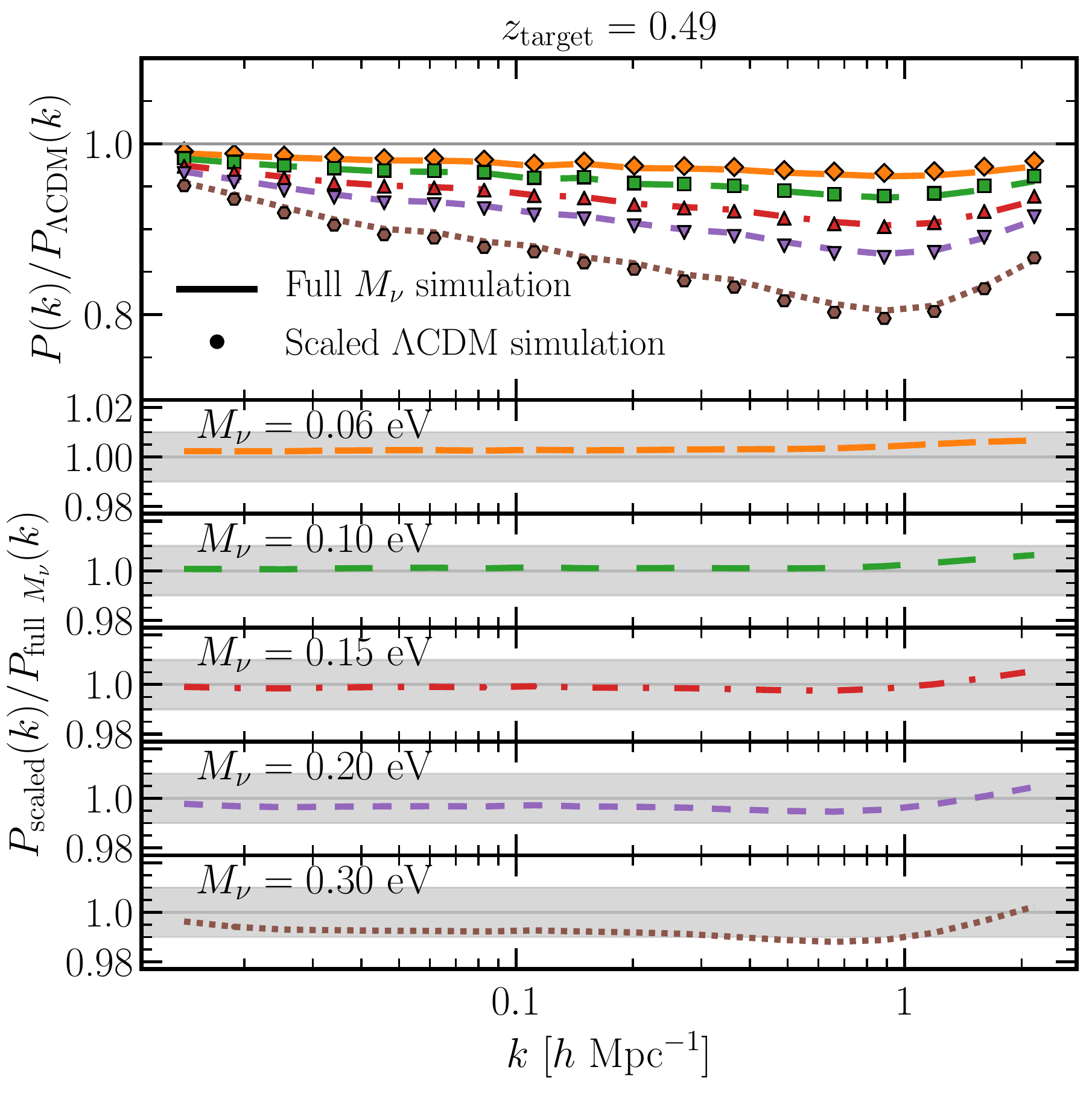}
  \includegraphics[width=0.48\textwidth]{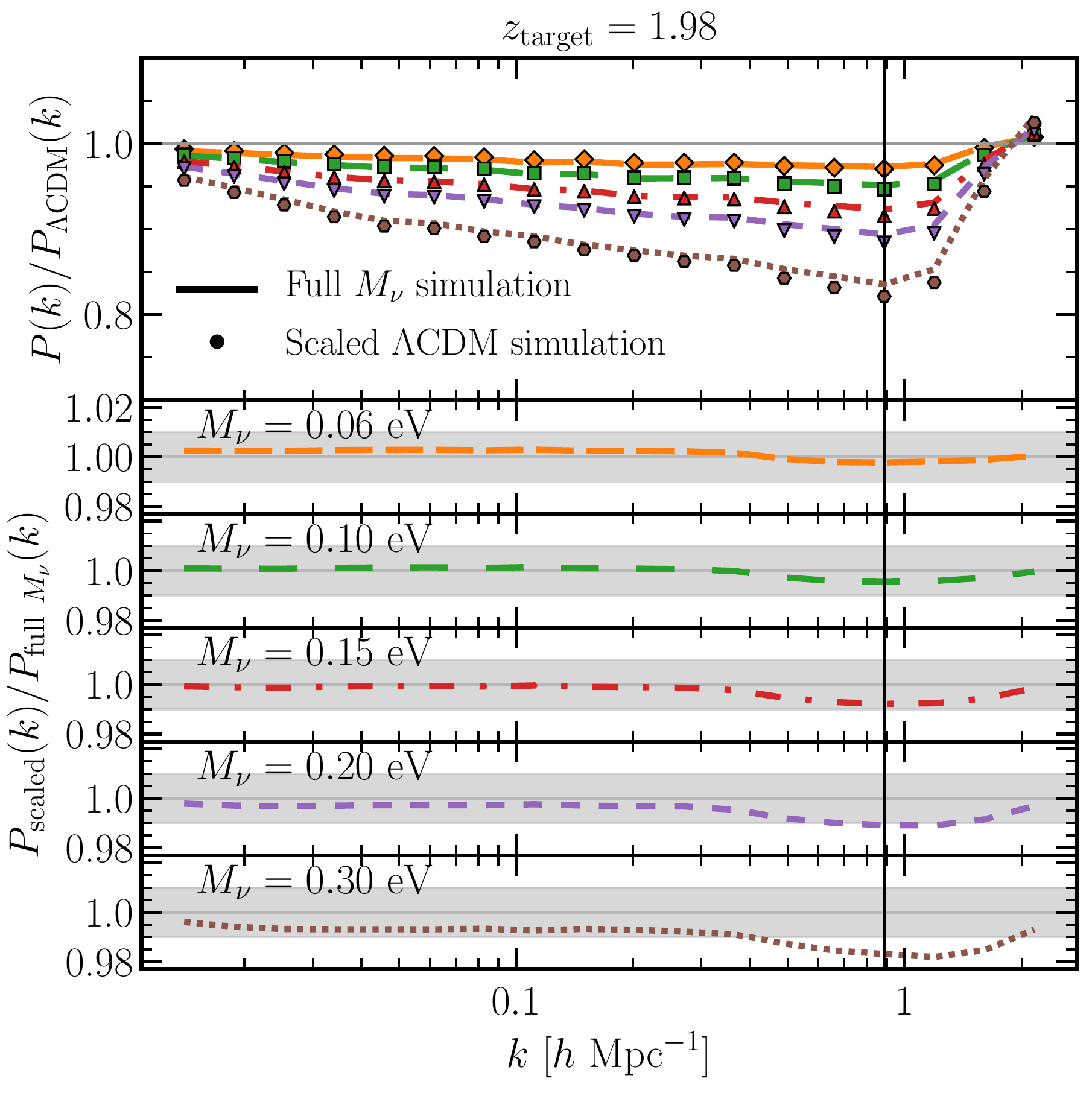}
  \caption{The accuracy of the scaling method tested against simulations run in the target cosmology, in terms of real space cold matter power spectrum. The original simulation is characterized by a pure $\Lambda$CDM cosmology without massive neutrino. The target cosmologies, instead, span the mass interval $M_\nu \in [0.06, 0.3]$ eV. We present two target redshifts, namely $z_{\rm t} \simeq 0.5$ and $z_{\rm t} \simeq 2$. \textit{Upper panel}: the lines show the ratio of the matter power spectrum measured in the simulations run including massive neutrinos and in the neutrinoless case; points show the same ratio computed using the $\Lambda$CDM simulations scaled to include massive neutrinos, divided by the $\Lambda$CDM simulation at the same redshift. \textit{Lower panels}: the ratio between the cold matter power spectrum measured in the scaled $\Lambda$CDM simulation and in the simulations actually run with neutrinos. The gray shaded region marks the 1\% level. The black vertical line marks the scale at which the shot noise starts accounting for more than 80\% of the signal.}
  \label{fig::sim-pk}
\end{figure*}

Structures in the simulations are identified employing a modified version of the \texttt{subfind} algorithm \citep{SpringelEtal2001} which we have inlined in \texttt{GADGET}. In order to define a Friend-of-Friends halo \citep{DavisEtal1985}, it must contain a minimum number of $N_{\rm min} = 20$ particles, with linking length $b = 0.2$ in units of the mean intraparticle distance. We also require that subhaloes contain at least $20$ particles.

\subsection{Matter power spectrum}

Fig. \ref{fig::sim-pk} shows the real-space matter power spectrum measured in the different neutrino simulations and in their corresponding $\Lambda$CDM scaled simulations. We consider the case in which we aim at reproducing neutrino cosmologies at $z \simeq 0.5$ and at $z \simeq 2$. Although not shown, we have checked that the results presented below also hold for redshift $z \simeq 1$.
All spectra presented here are measured on regular meshes characterised by a grid-size of $N_{\rm grid} = 512^3$ points. This corresponds to considering a range of wavemodes spanning from the fundamental model of the grid, $k_{\rm F} \simeq 9 \times 10^{-3}  ~ h ~ \mathrm{Mpc}^{-1}$, to the Nyquist mode $k_{\rm N} \simeq 2.3 ~ h ~ \mathrm{Mpc}^{-1}$. We employ two interlaced grids in order to suppress the contribution of aliasing arising from Fast Fourier Transform methods and we assign mass particles to the grids following a Triangular Shape Cloud (TSC) algorithm \citep[see, for example,][for an exhaustive treatment of accurate power spectrum estimators and their sources of noise]{SefusattiEtal2016}.
We stress that at $z \simeq 0.5$ the shotnoise accounts for $\sim 30 \%$ of the power spectrum at $k = 2 ~ h ~ \mathrm{Mpc}^{-1}$, while at $z \simeq 2$, at the same scale, it is the dominant contribution, accounting for $\sim 90 \%$ of the spectrum itself.

In the scaled simulation, as detailed in Sec. \ref{sec::method}, besides applying our transformation of lengths and times, we have also corrected large scales according to the linear theory of the target cosmology, and the small scales according to the halo-concentration correction. In the case presented in this figure, we have used the concentration-mass relation of \cite{LudlowEtal2016} (in Appendix \ref{appendix:concentration}, we test the impact of choosing other descriptions).

The ratio between the matter power spectrum measured in the $\Lambda$CDM scaled simulations and in the target neutrino simulations is compatible with $1$, with sub-percent accuracy for all neutrino masses considered, at all scales $k<2 ~ h ~ \mathrm{Mpc}^{-1}$. We highlight that, even though the amplitude and location of the baryonic acoustic oscillations are expected to vary with cosmology and after our length transformation, the ratios show no evidence of any oscillatory feature, confirming the accuracy of our large-scale correction.

Although always within $1\%$, the accuracy of the scaling somehwat degrades with neutrino mass, especially on scales approaching $k\simeq 1 ~ h ~ \mathrm{Mpc}^{-1}$. This originates from neglecting the impact of neutrinos on the concentration-mass relation and on the outer profiles of dark matter halos, aspects which could be improved in the future. Nevertheless, we stress that this level of disagreement is smaller than that induced by numerical and algorithmic choices in $N$-body simulations or than that induced by baryonic physics. For instance, already in the pure $\Lambda$CDM context, different $N$-body codes exhibit a disagreement of the low-redshift matter power spectra at the 1\% level at $k = 1 ~ h ~ \mathrm{Mpc}^{-1}$ \citep{SchneiderEtal2016}. Additionally, baryonic processes such as feedback from massive black holes, gas cooling, and star formation induce modifications to the mass power spectrum at the $10\%$ level for $k > 0.5 h ~ \mathrm{Mpc}^{-1}$. As a consequence, our rescaling procedure appears to be sufficiently accurate for modelling the galaxy population, given the intrinsic uncertainties and approximations usually made in $N$-body simulations.

\subsection{Halo abundance}

\begin{figure}
  \includegraphics[width=0.48\textwidth]{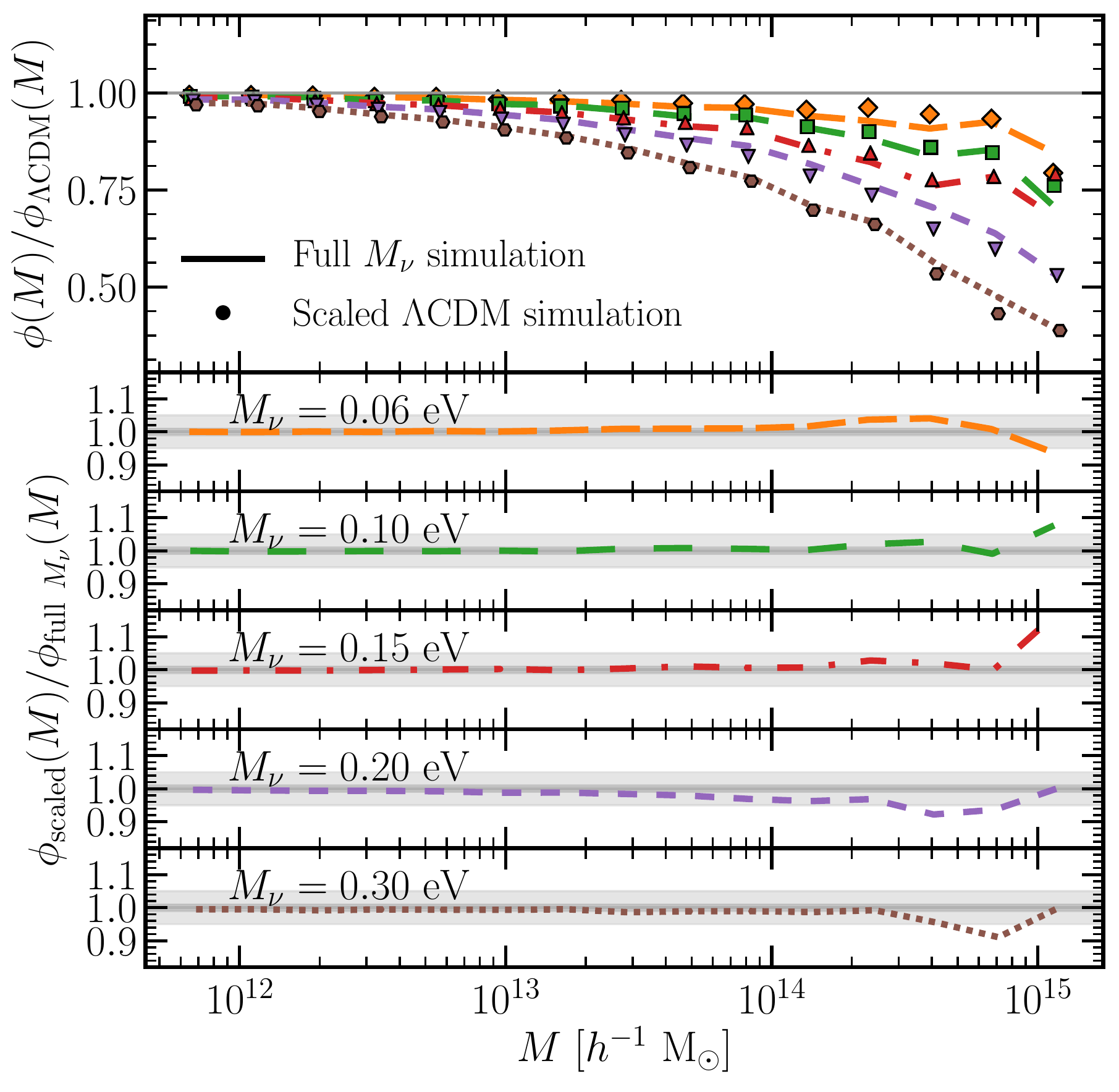}
  \caption{The accuracy of the scaling method for the halo abundance, tested against simulations actually run with massive neutrinos at the target redshift $z_{\rm t} \simeq 0.5$. \textit{Upper panel}: lines show the halo mass function measured in the simulations run in the five neutrino cosmologies, divided by the one measured in the $\Lambda$CDM simulation; points correspond to the ratio between the halo mass function measured in the scaled $\Lambda$CDM simulation and the one measured in the $\Lambda$CDM simulation at the same redshift. \textit{Lower panels}: the ratio between the halo mass function measured in the scaled $\Lambda$CDM simulation and the one measured in the corresponding neutrino simulation. The grey shaded region marks the 5\% level.}
  \label{fig::sim-hsmf}
\end{figure}

Since we based our scaling algorithm on the matching of the linear amplitudes of our $\Lambda$CDM cosmology and neutrino cosmology, we expect to find a very good agreement between the halo mass functions of the scaled $\Lambda$CDM simulation and the simulation run with neutrinos. In Fig. \ref{fig::sim-hsmf} we show that we achieve such agreement with $1\%$ accuracy for halo masses $M_{200} \lesssim 10^{14} ~ h^{-1} ~ \mathrm{M}_\odot$. Larger masses exhibit more scatter, which is within $5\%$ up to $M_{200} \lesssim 10^{15} ~ h^{-1} ~ \mathrm{M}_\odot$, and reaches $10\%$ for $M_{200} \gtrsim 10^{15} ~ h^{-1} ~ \mathrm{M}_\odot$.
However, due to the sparseness of the sample, the behaviour of the high-mass bins is dominated by Poisson noise, which in this case makes up more than 60\% of the signal.

\begin{figure}
  \includegraphics[width=0.48\textwidth]{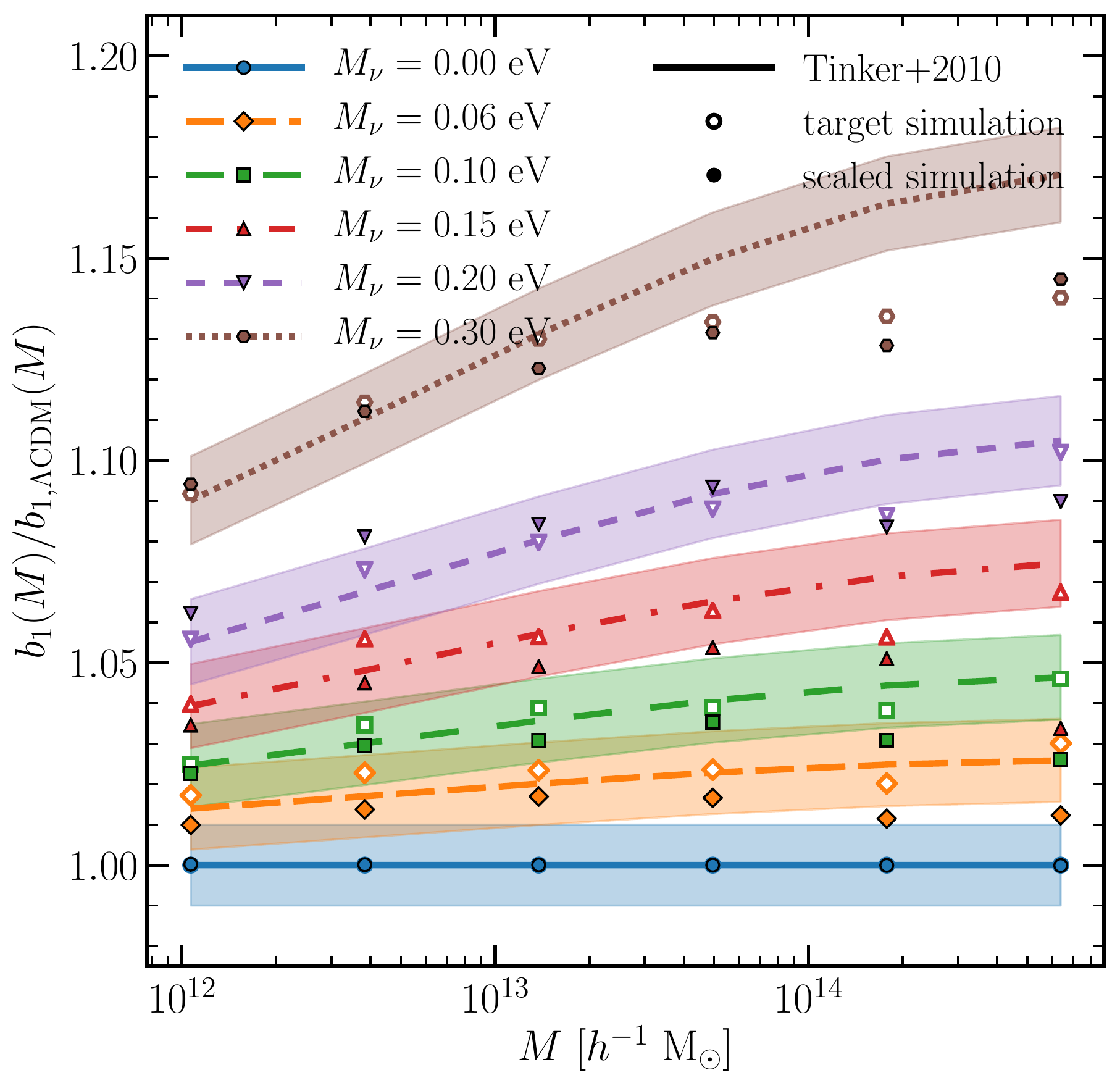}
  \caption{The linear bias parameter, obtained from the cross-power spectrum of the halo and matter fields, $b_1 = \sqrt{P_{hc} / P_{cc}}$, dividing the halo population in different mass bins at redshift $z \simeq 0.5$. Different colours correspond to different neutrino masses, in the usual range $M_\nu = \{0,0.06,0.10,0.15,0.20,0.30\}$ eV. Lines show the predicted bias-mass relation in each cosmology. Open markers refer to the bias-mass relation measured in the target simulations, while filled markers to that measured in the scaled simulations. The shaded areas mark a 1\% accuracy level around the theoretical predictions.}
  \label{fig::sim-bias}
\end{figure}

As an additional test of our rescaling procedure we investigate whether we are able to reproduce the linear halo bias in the presence of neutrinos. We present in Fig. \ref{fig::sim-bias} the linear bias parameter obtained from the cross-power spectrum of the halo and matter fields,
\begin{equation}
  b_1(M_{\rm h}) = \left\langle \left. \sqrt{\dfrac{P_{hc}}{P_{cc}}} \right|_{M_{\rm h}\in[M,M+\Delta M)} \right\rangle,
\end{equation}
where $\langle \cdot \rangle$ here denotes an average over the wavemode range $\kF < k < 0.1 ~ h ~ \mathrm{Mpc}^{-1}$, $\kF$ being the fundamental mode of the box. On this range, the linear bias is roughly constant, also in cosmologies including massive neutrinos (provided that it is defined with respect to the cold component, instead of the total matter field). In fact, slight deviations from scale independence have been observed \citep{ChiangEtal2018}, but are not considered here.

We find that our scaled simulations reproduce the bias mass relations of the simulations actually run including massive neutrinos with a 1\% accuracy in all mass bins. We also present in the same figure the predictions for the bias-mass relation obtained with the model presented in \cite{TinkerEtal2010}, applied, in the case with massive neutrinos, to the cold mass component. We notice that the accuracy of the rescaled simulations is actually higher than the agreement between the model and the measurements, especially in the high mass bins. This is competitive with the results of BE-HaPPY \citep{ValcinEtal2019}, the recently proposed bias emulator that also include a massive neutrino component. However, when the full minimization described in Sec. \ref{sec::method} is performed, our method guarantees the matching of the linear variance on a range of scales (i.e. the scales relevant to halo formation), thus avoiding the potential $\sigma_{8}$ mismatch discussed for this bias emulator.

\subsection{Galaxy clustering}

\begin{figure*}
  \includegraphics[width=\textwidth]{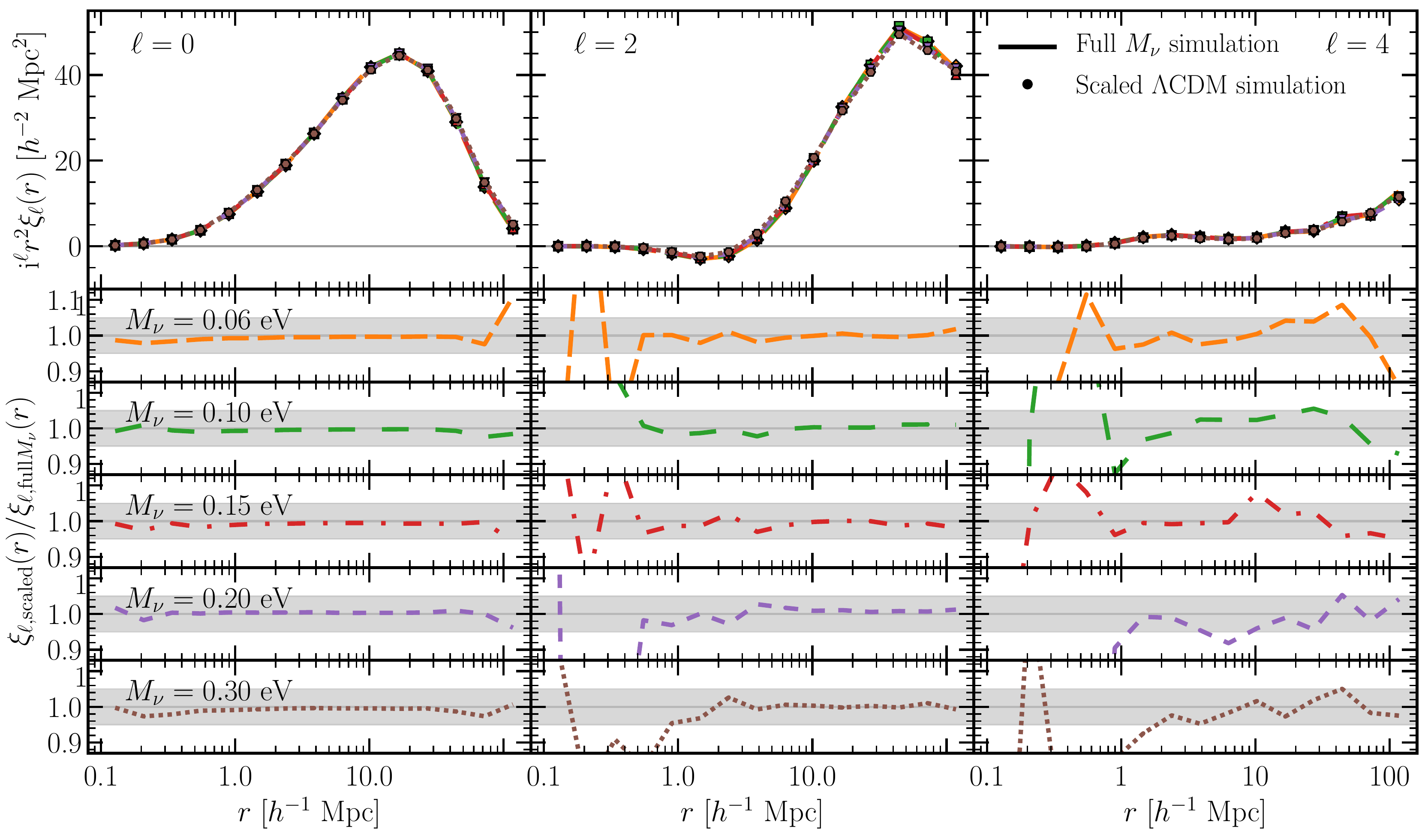}
  \caption{The accuracy of the scaling method for the monopole (left) and quadrupole (centre) and hexadecapole (right) of the galaxy 2-point correlation function in redshift space, tested against simulations run in the target cosmology. All cosmology share a fixed number density $n = 5 \times 10^{-3} ~ h^3~ \mathrm{Mpc}^{-3}$. \textit{Upper panel}: lines represent the 2-point correlation function multipoles measured in the simulations actually run including massive neutrinos; points correspond to the same quantities measured in the $\Lambda$CDM simulation, scaled to include neutrinos. \textit{Lower panels}: the ratio between the 2-point correlation function multipoles measured in the scaled $\Lambda$CDM simulations and the corresponding simulations actually run with neutrinos. The grey shaded region marks the 5\% level.}
  \label{fig::sim-mock}
\end{figure*}

Finally, we investigate the accuracy of the rescaling procedure in terms of galaxy clustering in redshift space. To build our galaxy sample we apply the subhalo abundance matching (SHAM) technique, assuming all galaxies reside in subhaloes, ranked according to one meaningful physical property \citep{ValeOstriker2004,GuoEtal2010}. In our case we select the maximum circular velocity of a subhalo, $v_{\rm max}$, as the physical property tracing the depth of the potential well each subhalo lives in \citep{Trujillo-GomezEtal2011}, neglecting any possible scatter. Please note that ranking according to $v_{\rm max}$ is not, in general, the optimal choice to reproduce realistic stellar mass functions or galaxy correlation functions \cite[see][for a extensive treatment of this matter]{Chaves-MonteroEtal2016}. However, high fidelity to observed data not being the main scope of this work, we use this galaxy sample as realistic enough to test the rescaling procedure.

Once rank-ordered according to $v_{\rm max}$, we retain a number of subhaloes such to obtain two samples, one with fixed galaxy number density of $n_{\rm g} = 5 \times 10^{-3} ~ h^3 ~ \mathrm{Mpc}^{-3}$ and the other with $n_{\rm g} = 5 \times 10^{-4} ~ h^3 ~ \mathrm{Mpc}^{-3}$, irrespective of cosmology. These number densities roughly correspond to those expected in JPAS and SphereX, and in EUCLID and DESI, respectively.

Fig. \ref{fig::sim-mock} shows the monopole, quadrupole, and hexadecapole of the 2-point redshift-space correlation function of our $n_{\rm g} = 5 \times 10^{-4} ~ h^3 ~ \mathrm{Mpc}^{-3}$ sample. We computed these correlation functions with the \texttt{Corrfunc} code \citep{SinhaGarrison2017}.
The upper panels show the multipoles measured both in the scaled $\Lambda$CDM simulations and in the target neutrino simulations, whereas the ratio between them is shown in bottom panels. Once again we find our method to perform at a high accuracy, and we are unable to find any sizeable bias given the statistical uncertainties of our simulations.

Specifically, typical disagreements are a few $\%$ level, which implies that any potential error introduced by the rescaling procedure is well below the error associated with the SHAM technique itself \citep[which is expected to reproduce the galaxy 2-point correlation function with an accuracy of $\sim 10\%$ for scales $R > 2 ~h^{-1}~\mathrm{Mpc}$, as shown in][]{Chaves-MonteroEtal2016}. Although not shown here, we found the same level of agreement for our sparser galaxy sample.

Noticeably, such accuracy is confirmed also for the quadrupole and hexadecapole of the galaxy correlation function, indicating that, while rescaling, the velocity field is properly reconstructed, both in terms of large-scale bulk motion, and of small scale dispersion inside clusters. As a consequence, we conclude that galaxy catalogues can be created for simulations rescaled in arbitrary neutrino cosmologies retaining all the desirable properties linked to the SHAM technique, i.e. its ability to reproduce galaxy velocity bias and assembly bias, while retaining all the internal properties of haloes.

Finally, it is interesting to note the similarity among the clustering for different neutrino masses, this despite the underlying differences in the halo mass function, bias, and halo occupation numbers. We will explore these effects in the future, but here we highlight the importance of modelling the galaxy population to correctly and accurately predict the observable signatures of neutrinos in the Universe.

\subsection{Non-gravitational couplings}

As pointed out in Sec. \ref{sec::method}, if a simulation is started by rescaling a $z=0$ matter power spectrum back to the initial time with a gravity-only growth factor, the matter power spectrum measured at high redshifts will show a mismatch on large scales with respect to the linear solutions of the Einstein-Boltzmann equations. This mismatch is partly due to the lack of residual couplings between the radiation and baryon distributions, owing to a non-zero Compton scattering and to the fact that, if we consider the Boltzmann solutions in the synchronous gauge, we are actually comparing quantities in different gauges.

An interesting use of our scaling algorithm is that, if desired, one can correct for these effects by including them as an additional contribution to the displacement field used in the large scale correction. To test this idea we have run a neutrino simulation (although the same applies to a $\Lambda$CDM simulation) that shares all characteristics with the simulations of our main suite, but includes $N_{\rm part} = 512^3$ particles in a comoving cube of side $L_{\rm box} = 2000 ~ h^{-1} ~ \mathrm{Mpc}$. This choice has been motivated by a balance between the amplitude of the effect (few percent at $k\simeq 10^{-3} ~ h ~ \mathrm{Mpc}^{-1}$ at $z=49$), and the expected shot noise.

\begin{figure}
  \includegraphics[width=0.48\textwidth]{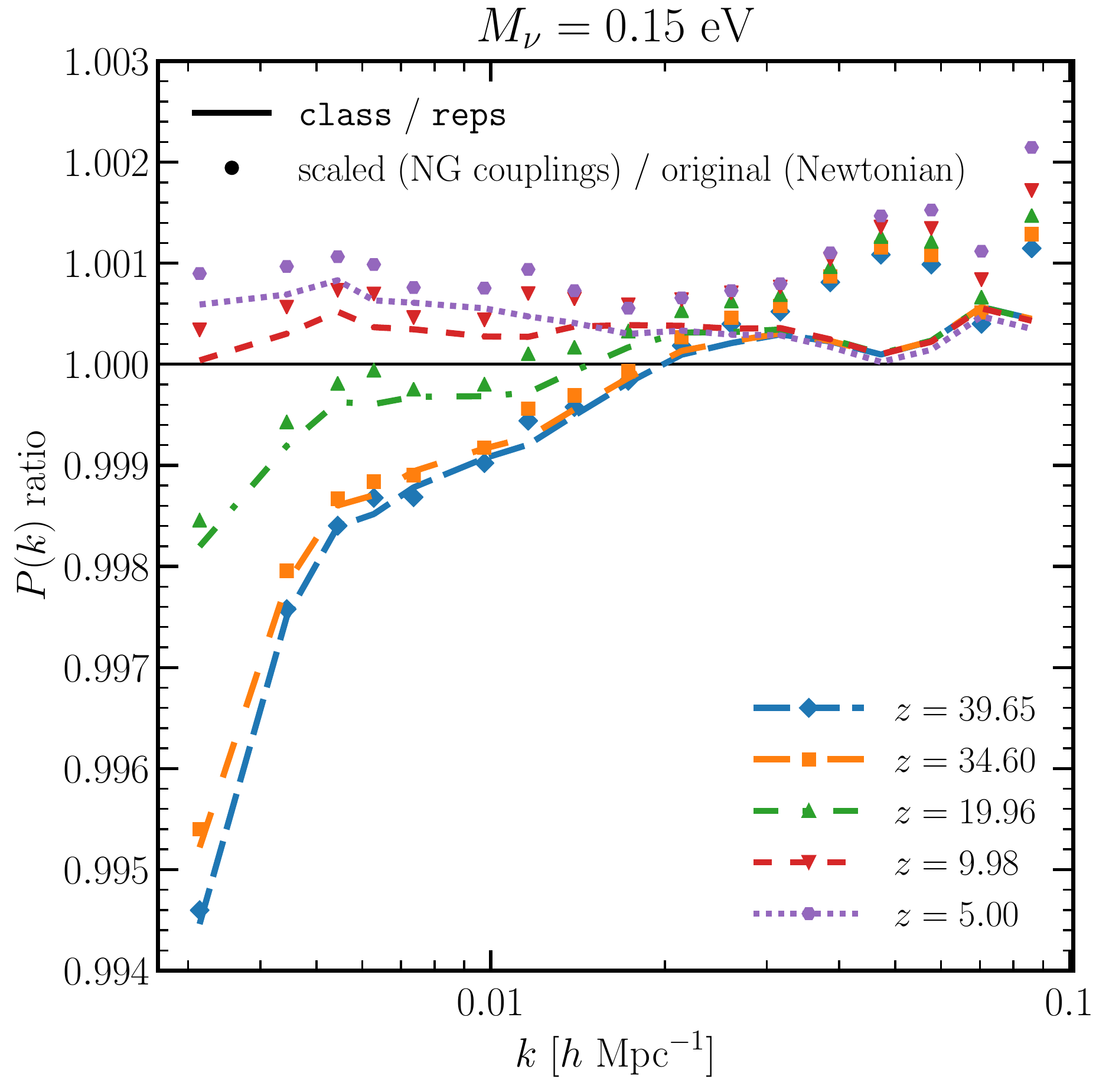}
  \caption{The implementation of a large scale correction that includes non-gravitational couplings absent in Newtonian N-body codes. Lines show, for different redshifts, the mismatch expected due to neglecting the effect of radiation perturbations on large scales; this is computed as the ratio between the relativistic solution obtained with \texttt{class} and the Newtonian backscaling of a $z=0$ power spectrum performed with \texttt{reps}. Point show the ratio between a simulation rescaled to include these relativistic effects and the original simulation, which does not include them.}
  \label{fig::relcorr}
\end{figure}

Fig. \ref{fig::relcorr} shows the ratio between the power spectrum obtained when adding the relativistic correction on large scales and the original one, corresponding to the purely Newtonian evolution of the initial conditions. We compare it with the prediction of the amplitude of this mismatch, computed taking the ratio of the relativistic solutions of \texttt{class} and the Newtonian rescaling of \texttt{reps}. We conclude that this method is able to reproduce the large scale clustering expected when solving the linearised Einstein-Boltzmann equations at high redshift. As expected, for $z<10$, this correction becomes superfluous, because the back-scaling method used to set-up the initial conditions of the simulations guarantees convergence at low redshift.

\subsection{Fitting formulae}

To the end of easing the computation of the length and time scaling parameters necessary for the scaling procedure, we present two empirical parametric fitting functions.
We find that the time and scale transformations mainly depend only on the total matter density parameter and on the neutrino density parameter, $\Omega_{\rm m,0}$ and $\Omega_{\nu,0}$ respectively. As in the rest of this work, we keep constant the total matter density parameter among the massive neutrino cosmology and the $\Lambda$CDM one.

As a result, in the case we want to add a neutrino component with $\Omega_{\nu,0} = M_\nu/(93.14 h^2)$ to a simulation run assuming a given $\Omega_{m,0}$ we will need to expand lengths by the factor given by
\begin{equation}
  s = \alpha ~ \Omega_{\nu,0} + \beta ~ \Omega_{\rm m, 0} + \gamma ~ \Omega_{\rm m,0} \Omega_{\nu,0} + \delta,
\end{equation}
where $\alpha, \beta, \gamma$ and $\delta$ are free parameters, whose bestfitting values we report in Table \ref{tab::nu-recipe}.
This result does not depend on the target time we desire to reproduce in the massive neutrino cosmology of choice.

Moreover, we also have to consider a earlier snapshot of our $\Lambda$CDM simulation. We employ the same functional form as for the length rescaling, with different parameters (namely $\varepsilon, \zeta, \eta, \vartheta$). However, in this case, which previous output to consider depends on the target time we want to reproduce in the neutrino cosmology,
\begin{equation}
  a^*(a_{\rm target}) = \varepsilon ~ \Omega_{\nu,0} + \zeta ~ \Omega_{\rm m, 0} + \eta ~ \Omega_{\rm m,0} \Omega_{\nu,0} + \vartheta,
\end{equation}
where $a^*$ and $a_{\rm target}$ are the expansion factor selected by the minimization procedure and the expansion factor at the target time respectively.
As a useful example, we present the fitting formula calibrated assuming four significant values of the target expansion factor, corresponding to the target redshifts $z_{\rm t} = \{0, 0.5, 1.0, 2.0\}$.
The results of the fitting procedures are presented in Table \ref{tab::nu-recipe}.

\begin{table}
  \begin{tabular}{l c c c c}
    \hline
    & $\alpha$ & $\beta$ & $\gamma$ & $\delta$ \\
    $z_{\rm t} = 0.0$ & $12.48094$ & $0.01508$ & $-0.24582$ & $0.99361$ \\
    \hline
    & $\varepsilon$ & $\zeta$ & $\eta$ & $\vartheta$ \\
    $z_{\rm t} = 0.0$ & $-45.20044$ & $0.11578$ & $0.81595$ & $0.95602$ \\
    $z_{\rm t} = 0.5$ & $-22.95795$ & $0.02504$ & $0.41368$ & $0.65882$ \\
    $z_{\rm t} = 1.0$ & $-14.82997$ & $0.00763$ & $0.26286$ & $0.49853$ \\
    $z_{\rm t} = 2.0$ & $-8.61413$ & $0.00118$ & $0.15002$ & $0.33387$ \\
    \hline
  \end{tabular}
  \caption{Fitted values of the parameters of the empirical functions that allow to quickly reproduce the length scaling parameter and the expansion factor in the original $\Lambda$CDM cosmology corresponding to a target cosmology with arbitrary neutrino mass. The function does not depend on the choice of target redshift for the length scaling parameter (an arbitrary value of $z_{\rm t} = 0$ is reported here), while it does for the expansion factor.}
  \label{tab::nu-recipe}
\end{table}

\begin{figure*}
  \includegraphics[width=\textwidth]{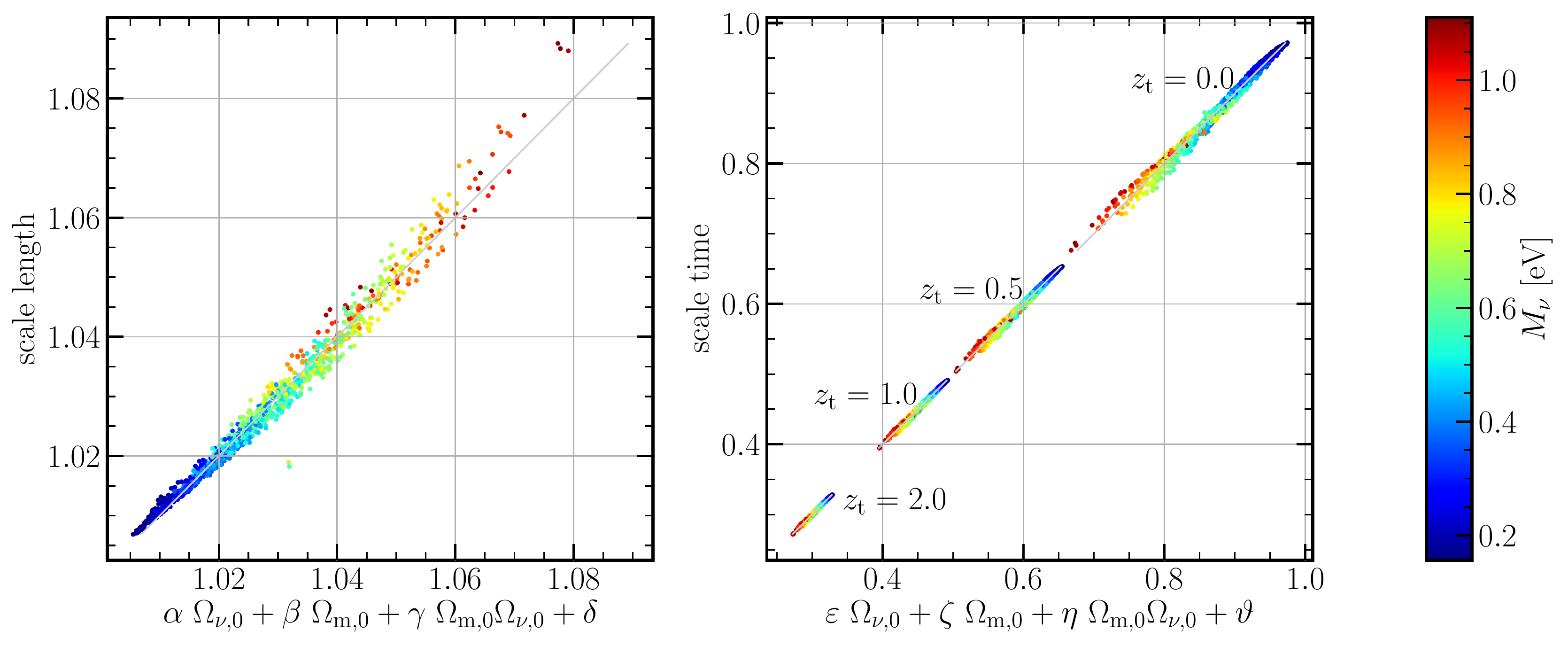}
  \caption{The dependence of the scaling parameters on cosmology, for the case in which the only cosmological parameter changed is the total neutrino mass. The larger the neutrino mass, the larger the simulation box size must become, following the relation shown in the left panel. This result does not depend on redshift. On the contrary, the time scaling does depend on the target redshift at which we want to reconstruct the neutrino cosmology. In general, given a redshift of the target neutrino cosmology, the corresponding $\Lambda$CDM simulation output will always be at an earlier time. In the specific case presented here, the target redshift considered are $z_{\rm t}=\{ 0, 0.5, 1, 2 \}$.}
  \label{fig::recipe}
\end{figure*}

\section{Conclusions}\label{sec::conclusion}

In this work, we have presented an extension of the cosmology scaling algorithm introduced in \cite{AnguloWhite2010} in order to include the effects of massive neutrinos. In particular, employing this method, it is possible to rescale a $\Lambda$CDM simulation, run without including massive neutrinos, to any neutrino cosmology.

The cosmology rescaling algorithm relies on the matching (on a range of scales) of the linear amplitude of fluctuations of the target cosmology at a given redshift with that of the original cosmology, at a redshift and with a length transformation to be determined. By matching the variance, in the Press-Schechter formalism, the halo abundance is automatically reconstructed as the one of the target cosmology. However, more care is needed to accurately reconstruct the clustering properties of the matter, halo and galaxy fields. In the general argument, in particular, one has to correct for different large-scale evolution (including the position of the BAO peak), for differences in the small-scale nonlinear velocities (due to the different growth history), and for different halo concentrations (due to differences in the formation time of haloes).

In the specific massive neutrino case, there are some additional aspects that need to be taken care of. First of all, the growth factor and growth rate of matter perturbations exhibit a peculiar scale dependence induced by the phenomenon of neutrino free streaming. Such scale dependence, moreover, evolves in amplitude and shape, being more pronounced at later redshifts. Secondly, the relevant quantity to be matched between $\Lambda$CDM and neutrino cosmology is no longer the total mass linear variance, but the cold matter linear variance. This is due to the neutrino free streaming scale being larger than typical halo sizes at all times, thus preventing neutrinos from playing any significant role in the context of halo collapse. Lastly, by correcting the large scale modes of the matter distribution, the correct neutrino scale dependence is recovered, with the expected, neutrino-mass dependent large-to-small scale ratio.

By comparing a $\Lambda$CDM simulation scaled to reproduce neutrino cosmologies with total masses $M_\nu = \{0.06,0.10,0.15,0.20,0.30\}$ eV to their corresponding simulations, run actually including massive neutrinos, we showed that we are able to reproduce the target matter power spectrum with 1\% accuracy on all scales $k<2~h~\mathrm{Mpc}^{-1}$, for all the choices of neutrino mass. Extending the comparison to halo abundance we verify that the halo mass function is reconstructed with better than 5\% accuracy.
Finally, also the redshift-space 2-point correlation function monopole and quadrupole of a galaxy population obtained with a SHAM technique are reconstructed in the target cosmology with a mismatch lower than 5\% with respect to the actual neutrino simulations.

This method is in principle applicable to any $\Lambda$CDM simulation, which is then converted into a neutrino simulation in a quick way and inheriting the, potentially high, accuracy.
As a consequence, the cosmology rescaling method opens up the possibility of performing analyses of cosmological datasets directly against simulations. As a matter of fact, considerable amount of information is encoded in the physical processes at play at the cluster or galaxy scale, where the evolution of the matter and galaxy fields is highly nonlinear. Besides improving models, another method is to consider the simulations themselves \textit{as the model}. Together with physically motivated galaxy formation models, this provides the most accurate predictions for LSS analyses of forthcoming surveys. Thanks to the work here presented, such analysis can also include (and constrain) the effect of massive neutrinos.

\section*{Acknowledgements}

We acknowledge the support of the Spanish MINECO and of the European Research Council through grant numbers  \textit{AYA2015-66211-C2-2} and \textit{ERC-StG/716151}.

%%%%%%%%%%%%%%%%%%%%%%%%%%%%%%%%%%%%%%%%%%%%%%%%%%

%%%%%%%%%%%%%%%%%%%% REFERENCES %%%%%%%%%%%%%%%%%%

% The best way to enter references is to use BibTeX:

\bibliographystyle{mnras}
\bibliography{Bibliography_all} % if your bibtex file is called example.bib

%%%%%%%%%%%%%%%%%%%%%%%%%%%%%%%%%%%%%%%%%%%%%%%%%%

%%%%%%%%%%%%%%%%% APPENDICES %%%%%%%%%%%%%%%%%%%%%

\appendix

% \section{Incorporating relativistic perturbations}

\section{Dependence on the choice of concentration-mass relation model} \label{appendix:concentration}

\begin{figure}
  \includegraphics[width=0.48\textwidth]{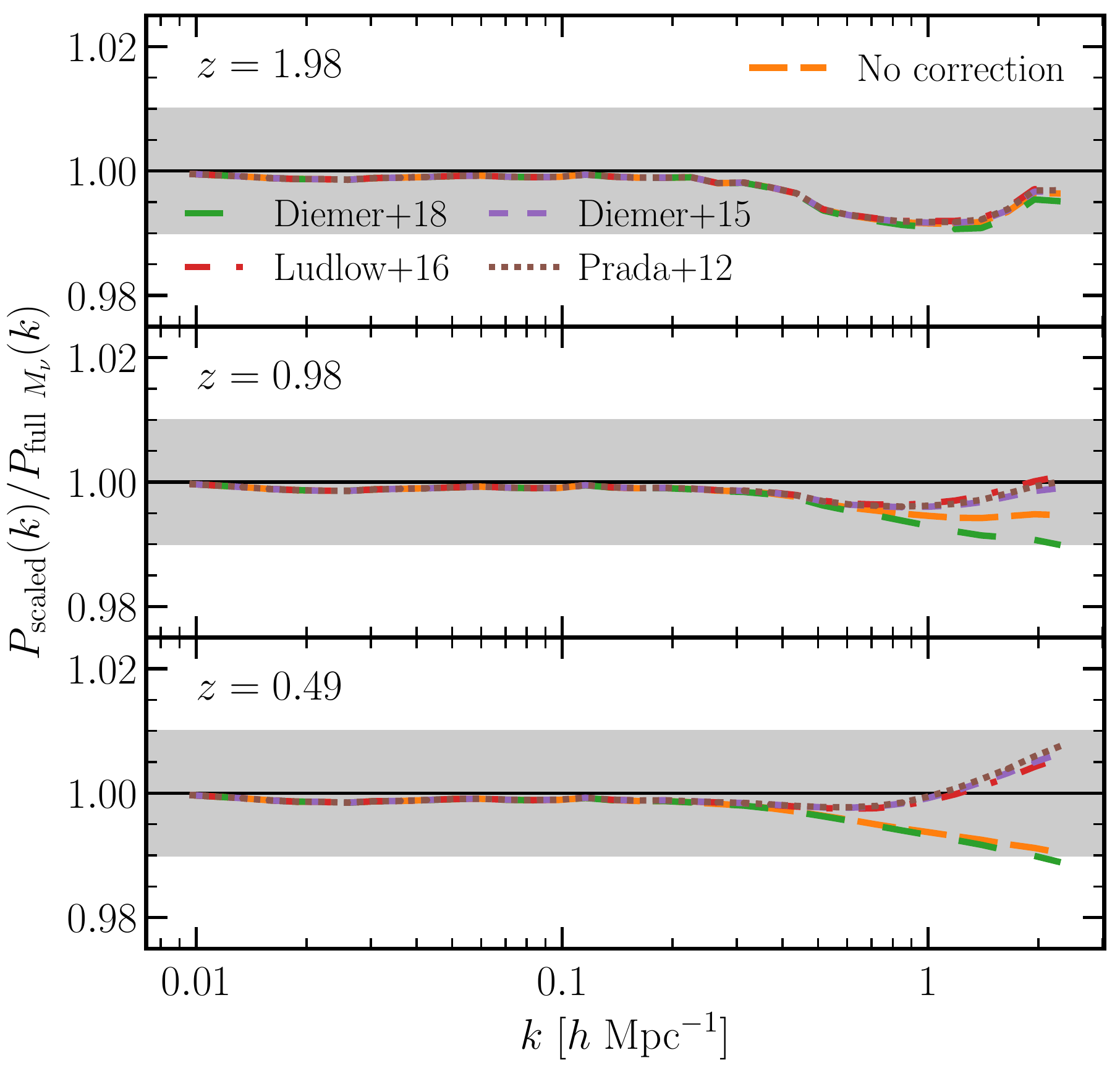}
  \caption{The difference in the cold matter power spectrum of the scaled simulation due to choosing different mass-concentration relation models to compute the concentration correction. We only show here the case of $M_\nu = 0.15$ eV, but similar results apply as well to the other neutrino masses considered in this work. Results are shown from top to bottom at three target redshifts, namely $z_{\rm t} \simeq \{2, 1, 0.5\}$. Orange dashed lines represent the case in which no concentration correction in applied, while the other lines correspond each to one of the different models quoted in the text.}
  \label{fig::conc-corrs}
\end{figure}

In Fig. \ref{fig::conc-corrs} we present a synoptic view of the performance of the rescaling technique at different redshifts for our mid-range neutrino mass, $M_\nu = 0.15$ eV. We stress, in particular, that we do not find significant differences when using different concentration-mass models to compute the concentration correction. Specifically, here we show the cases obtained using the models by \cite{DiemerMichael2018},  \cite{LudlowEtal2016}, \cite{DiemerKravtsov2015}, and \cite{PradaEtal2012}, finding that differences are way sub-percent.

%%%%%%%%%%%%%%%%%%%%%%%%%%%%%%%%%%%%%%%%%%%%%%%%%%

% Don't change these lines
\bsp	% typesetting comment
\label{lastpage}
\end{document}